\documentclass[aps,prl,twocolumn,amsmath,amssymb,showpacs,longbibliography]{revtex4-1}

\usepackage{ifpdf} 
\usepackage[sort&compress]{natbib}
\usepackage{color}
\usepackage{soul}
\usepackage{dcolumn}
\usepackage{commath}
\usepackage{amsmath,latexsym}
\ifx\pdfoutput\undefined 
\usepackage{graphicx} 
\else 
\usepackage[pdftex]{graphicx} 
\usepackage{epstopdf} 
\fi 
\usepackage[normalem]{ulem}

\preprint{}

\begin{document}
\title{RFOT theory for glassy dynamics in a single condensed polymer} 

\author{Hyun Woo Cho$^{1}$, Guang Shi$^{1}$, T. R. Kirkpatrick$^{2}$ and D. Thirumalai$^{1}$} 
\affiliation{$^{1}$Department of Chemistry, University of Texas at Austin, Austin, Texas 78712, USA}
\affiliation{$^{2}$Institute for Physical Science and Technology, University of Maryland, College Park, Maryland 20742, USA}

\begin{abstract}
The number of compact structures of a single condensed polymer (SCP), with similar free energies, grows exponentially with the degree of polymerization. In analogy with structural glasses (SGs), we expect that at low temperatures chain relaxation should occur by activated transitions between the compact metastable states.     By evolving the states of the SCP that is linearly coupled to a reference state, we show that, below a dynamical transition temperature ($T_d$), the SCP is trapped in a metastable state leading to slow dynamics.   At a lower temperature, $T_K \ne 0$, the configurational entropy vanishes, resulting in a thermodynamic random first order ideal glass transition.  The relaxation time obeys the Vogel-Fulcher-Tamman law, diverging at $T=T_0 \approx T_K$. These findings, accord well with the random first order transition theory, establishing that SCP and SG exhibit similar universal characteristics. 


\end{abstract}

\maketitle

The discovery that a single chromosome (long mega base pair sized polymers) is heterogeneously organized \cite{Finn19Science} has triggered a keen interest in its dynamics. Experiments have suggested that glass-like behavior should be expected  in chromosome dynamics \cite{Bronshtein15NatComm}, translocation of DNA \cite{Berndsen8345,Keller:2016bh}, collapse kinetics of polymers \cite{PhysRevLett.119.087801}, and more recently intrinsically disordered proteins \cite{Morgan20PRL}.  Dynamics \cite{Perkins822,C1SM05298E,PhysRevLett.114.178102}, and phase behavior  \cite{PhysRevLett.76.3029,Rampf2006,Taylor2009} of single (synthetic) polymers have also suggested that a single polymer displays glassy behavior upon cooling or compression, just like in bulk liquids that undergo a transition to glassy states \cite{PhysRevE.65.030801,Tress1371,macr6b01461}.  However, it is unknown whether the glass transition in a single polymer, which is relevant in a number of unrelated systems,  is governed by the same physical principles that describe their macroscopic counterparts.  

Single condensed polymer (SCP) should exhibit glass like behavior because their phase space structure satisfies all the requirements for observing slow dynamics. 
At  temperatures ($T$s) below the collapse temperature, $T_\theta$, the number of compact structures or states, with similar free energies, scales exponentially with $N$ \cite{Orland85JPhys}. The states are separated by  large free energy barriers.  At high temperatures, the relaxation dynamics is expected to be diffusive. At low temperatures, the exploration of the distinct compact structures can only occur by activated transitions. Thus,  the  physical picture for a SCP is exactly the same as in the structural glass transition (SGT). Therefore, we expect that the dynamics of the SCP should be described by theories developed for bulk glassy systems. In particular, we anticipate that the SCP dynamics, over a wide range of temperatures, can be understood within the framework of the Random First Order Transition (RFOT) theory.  That this is so is the main conclusion of this work.  

\begin {figure*}
\centering
\includegraphics  [width=6.in] {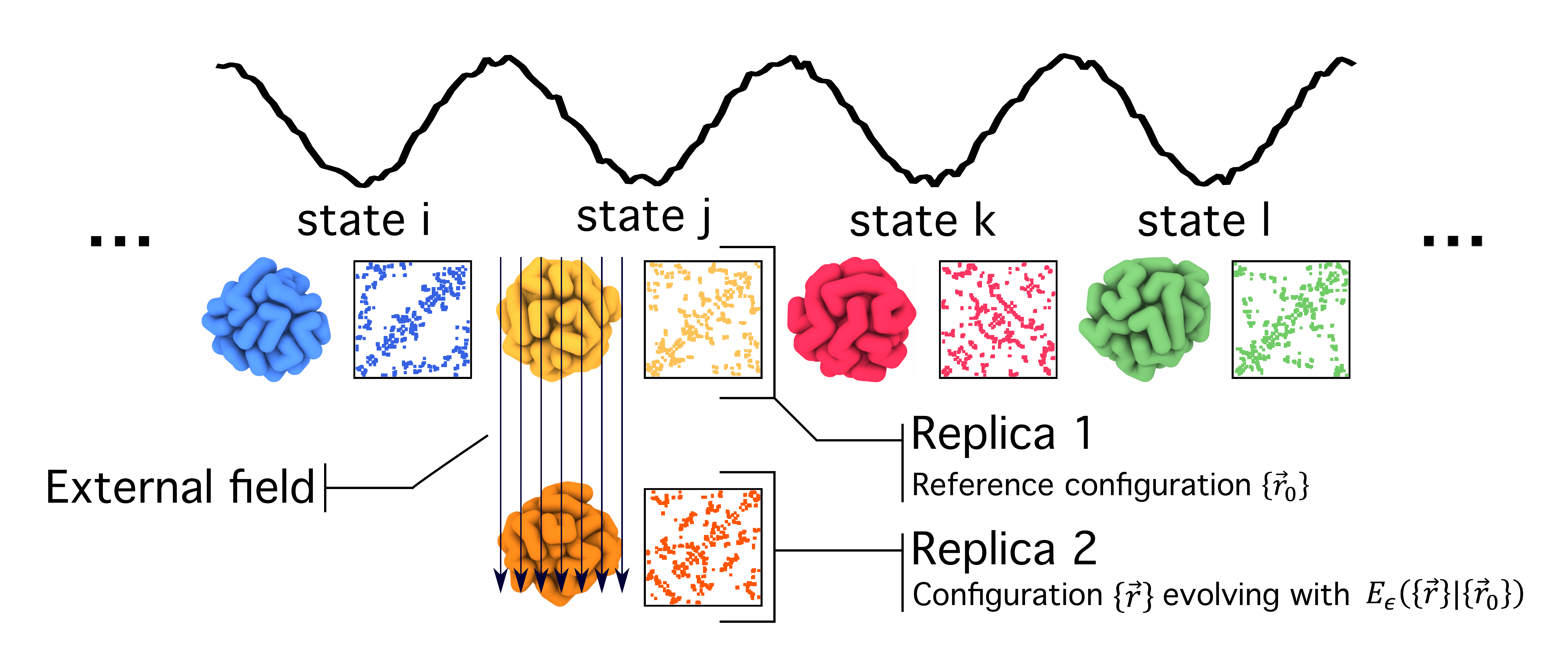}
\caption{\textbf{Numerical implementation of the Franz-Parisi (FP) method for polymers.} In the FP method, two replicas (Replica 1 and Replica 2) are coupled. The quenched reference conformation in Replica 1 is sampled from an ensemble of equilibrium SCP conformations.   The conformation in Replica 1 represents one of the  exponentially large number of metastable sates. The conformation in Replica 2, coupled to Replica 1 by an external field, is evolved using the Monte Carlo method. The panels next to the snapshot represent the contact maps in the metastable states. The structural similarity between Replica 1 and Replica 2 is defined using the extent of overlap of the contact maps (blue and red for example).}
\label{schematic}
\end{figure*}

Let us describe the salient aspects of the energy landscapes of SCP and liquids that undergo the SGT.  The dynamics in the latter case is well-described by the Random First Order Transition  (RFOT) theory \cite{PhysRevA.40.1045}, developed based  on precisely solvable models  \cite{Kirkpatrick87PRL,Kirkpatrick87PRB,Kirkpatrick-Wolynes87PRB}. An important physical ingredient  in the RFOT theory for the SGT  is the emergence of an exponentially large number of metastable states with similar free energies \cite{Goldstein69JCP} below a characteristic dynamic transition temperature $T_d$ \cite{Kirkpatrick87PRL,Kirkpatrick87PRB}.  At $T < T_d$, the system is trapped in a specific metastable state for a finite but long time period, and hence relaxation occurs only by activated transition.  The free energy barrier, $\Delta F^\ddagger$, between the metastable states is related to the configurational entropy, $S_{\text{conf}}$ as $\Delta F^\ddagger\sim S_{\text{conf}}^{-1}$ \cite{PhysRevA.40.1045}. Since $S_{\text{conf}}$ decreases as $T$ decreases, $\Delta F^\ddagger$ increases, resulting in a significant increase in the structural relaxation time.  RFOT theory predicts that $S_{\text{conf}}$ vanishes at an ideal glass transition temperature, $T_K<T_d$, at which a thermodynamic random first-order transition, without latent heat, from a super cooled liquid to an ideal glass occurs. 

To affirm the predictions of the RFOT theory in the SCP, we use the Franz-Parisi (FP) method \cite{PhysRevLett.79.2486,Franz_2013}, which involves coupling two copies of the system through a field with strength $\epsilon$.  FP showed that an order parameter measuring the structural similarity between the states exhibited first order transition at non-zero $\epsilon$ only when a large number of metastable states emerge.   
The FP method was developed further to establish the emergence of metastable states, and possible the existence of an ideal glass transition in model atomic glasses \cite{PhysRevE.88.022313,PhysRevE.89.022309,PhysRevLett.114.205701,Berthier11668,Berthier11356}. 


We used a standard bead-spring model for a polymer composed of a sequence of $N=128$ Lennard-Jones (LJ) particles linearly connected by a harmonic potential {(section I of the Supplementary Information (SI)). At low temperatures, the polymer is trapped in a metastable state instead of transitioning to an ordered state (section II in the SI). Following FP (Figure~\ref{schematic}), we created two replicas (or samples) of the SCP at the same $T$. Replica 1 in Figure~\ref{schematic} represents a fixed reference conformation ($\{\vec{r}_0\}$), chosen from an equilibrium ensemble, and serves as a quenched random field. The conformation of  $\{\vec{r}\}$ in replica Replica 2 (Figure~\ref{schematic}) is evolved using Monte Carlo simulation by coupling it to Replica 1 through an external  field (External field in Figure~\ref{schematic}). The energy $E_\epsilon(\{\vec{r}\}|\{\vec{r}_0\})$ of the coupled replicas is, 
\begin{eqnarray}
E_\epsilon(\{\vec{r}\}|\{\vec{r}_0\})=E(\{\vec{r}\})-N\epsilon Q_{\text{stat}}(\{\vec{r}\},\{\vec{r}_0\}),
\label{Hamil}
\end{eqnarray}
where $E(\{\vec{r}\})$ is the potential energy of the SCP in Replica 2, $\epsilon$ is the strength of the external field, and $Q_{\text{stat}}(\{\vec{r}\},\{\vec{r}_0\})$ measues the structural similarity between $\{\vec{r}\}$ and $\{\vec{r}_0\}$. 


By anticipating applications to chromosomes, we used contact maps, which can be inferred from cross-linking experiments \cite{Finn19Science}, to calculate $Q_{\text{stat}}$. The contact map is a low dimensional representation of given structure of the SCP,  as are the Hi-C maps for the chromosomes.   
Two monomers that are not covalently linked to each other are in contact if the distance between them is less than $R_\text{c} = 1.4\sigma$, corresponding to the first minimum in the radial distribution function ( section II in the SI). Panels next to the snapshots  in Figure~\ref{schematic} illustrate examples of the contact maps, where the pairs of non-bonded monomers in contact are projected onto a two dimensional space. 

The static overlap function $Q_\text{stat}$  is calculated using,
\begin{eqnarray}
Q_{stat}(\{\vec{r}\},\{\vec{r}_0\})=\frac{\sum _{(i,j)'}q_{ij}(\{\vec{r}\})q_{ij}(\{\vec{r}_0\})}{\sum_{(i,j)'}q_{ij}(\{\vec{r}_0\})}, 
\label{Order}
\end{eqnarray}
where $q_{ij}(\{\vec{r}\})$ is the contact function of configuration $\{\vec{r}\}$, which is unity if $i$ and $j$ monomers are in contact and zero otherwise, and $(i,j)'$ indicates that the sum is over all non-bonded pairs of monomers. If  $\{\vec{r}\}$ and $\{\vec{r}_0\}$ are identical, $Q_{\text{stat}}(\{\vec{r}\},\{\vec{r}\})$=1. On the other hand, if the replicas are totally uncorrelated, $Q_{\text{stat}}$ is the average contact probability $\langle q_{ij} \rangle$, which is found to be $\simeq0.13$ for the parameters used in the simulations. From Eqs~(\ref{Hamil}) and (\ref{Order}), it follows that $Q_{\text{stat}}$ varies as a function of $\epsilon$. When the external field strength, $\epsilon$, is sufficiently large, $\{\vec{r}\}$ is highly biased to $\{\vec{r}_0\}$, such that $Q_{\text{stat}}\simeq1$. If $\epsilon$ decreases to 0, $\{\vec{r}\}$ is independent of $\{\vec{r}_0\}$, resulting in $Q_{\text{stat}}\simeq\langle q_{ij} \rangle$. 

It is important to note that  the expected changes in $Q_{\text{stat}}$ with $\epsilon \ne 0$ should have the characteristics of first order transition only if metastable states are probed.  The randomly-chosen equilibrium configurations in Replica 1 is an example of a particular metastable state (Figure~\ref{schematic}). Since all the metastable states are generated at  the same thermodynamic condition (the same $T$ and $N$), they are equivalent. The individual free energy is $F_\alpha$ where $\alpha$ is a label for the metastable state.  The canonical free energy $F_{\text{tot}}$ is less than the component averaged free energy, $\sum_{\alpha} P_{\alpha} F_\alpha$, where $P_{\alpha}$ is the probability of being in the state $\alpha$ \cite{Palmer82AdvPhys,PhysRevA.40.1045,RevModPhys.87.183}. The difference between the two is $TS_{\text{conf}}$, which is the the entropic gain arising from an exploration of all possible states, $F_{\text{tot}}=F_\alpha-TS_{\text{conf}}$. Thus, if $\epsilon$ is strong enough to compensate for the entropic penalty, the SCP in Replica 2 would be trapped in a single metastable state of the SCP, such that $Q_{\text{stat}}=Q_{\text{glass}}\simeq1$. Otherwise, it could explore all possible metastable states over time, resulting in $Q_{\text{stat}}$ being equal to $Q_{\text{liquid}}=\langle q_{ij} \rangle$. This argument shows that, at the critical value of $\epsilon=\epsilon_c$ where $N\epsilon_c (Q_{\text{glass}}-Q_{\text{liquid}}) \simeq TS_{\text{conf}}$, $Q_{\text{stat}}$ should change discontinuously between $Q_{\text{glass}}$ and $Q_{\text{liquid}}$, which would be a signature of a first order transition. Thus, by showing $Q_{\text{stat}}$ exhibits the first-order-like transition at  $\epsilon \ne 0$, we can confirm the existence of metastable states in the SCP.  

\begin {figure*}
\centering
\includegraphics  [width=6.in] {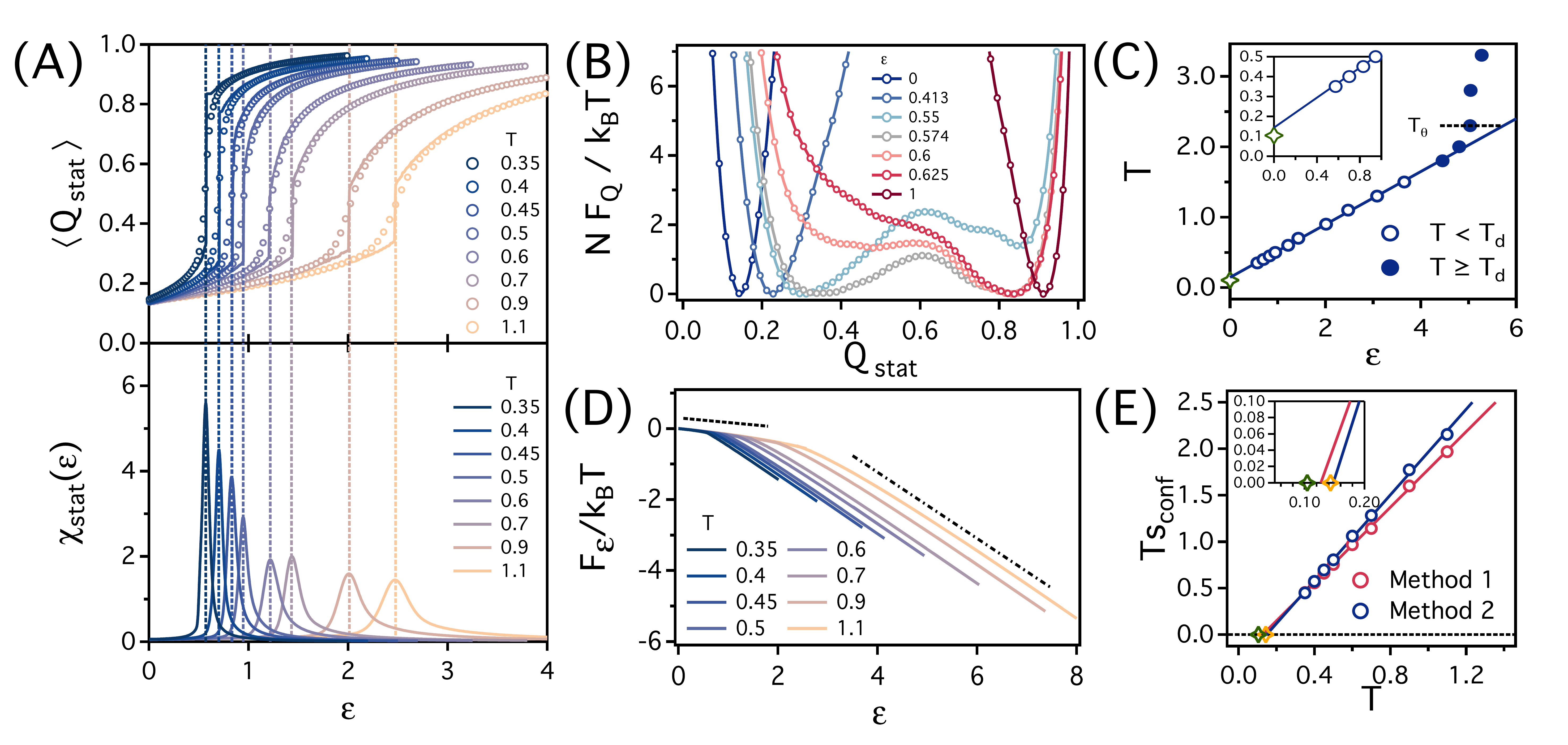}
\caption{\textbf{Phase behavior of $Q_{\text{stat}}$.} (A) Average $\langle Q_{\text{stat}} \rangle$ (the upper panel) and susceptibility $\chi_{\text{stat}}(\epsilon)$ (the lower panel) of the order parameters with respect to $\epsilon$ for various $T$. The vertical dashed lines represent the position of $\epsilon$ where $\chi_{\text{stat}}(\epsilon)$ has the maximum value. The solid lines represent the position of $Q_{\text{stat}}$ where $F_Q$ in (B) has the minimum value. (B) $F_Q$ of various $\epsilon$ at $T$=0.35. $F_Q$ in the graph is shifted by its minimum value. (C) $T-\epsilon$ phase diagram of $Q_{\text{stat}}$. The open and filled symbols represent $T$ of $\epsilon_c$ for $T<T_d$ and $T\geq T_d$, respectively. The solid line is a linear fit for $T<T_d$. The black dashed line denotes the position of $T_\theta$. The open green star denotes $T_0$. The inset is the same phase diagram at $0\leq \epsilon\leq0.9$. (D) $F_\epsilon$ as a function of $\epsilon$ for various $T$. $F_\epsilon$ is shifted by $F_{\epsilon=0}$. The slopes of the dashed and dashed-dotted lines are $-0.13$ and $-0.93$, respectively. (E) $Ts_{\text{conf}}$ of single polymers as a function of $T$. The solid lines are their linear fits. The yellow and green open symbols represent $T_0$ and $y$ intercept of the solid line in (C), respectively. In the inset, we magnify the graph near the $x$ intercepts of the linear fits.}
\label{phase}
\end{figure*}

In the upper panel of Figure~\ref{phase} (A), $\langle Q_{\text{stat}}\rangle$ is plotted as a function of $\epsilon$ for various $T$ (the open symbols), where $\langle\cdots\rangle$ is a double average over thermal and disorder fluctuation; we first perform an average of $Q_{\text{stat}}$ over thermal fluctuations with $\{\vec{r}_0\}$ fixed, and then an average over various $\{\vec{r}_0\}$ is calculated to account for disorder fluctuations (details are in section II of the SI). As expected, $\langle Q_{\text{stat}}\rangle$ decreases from $\simeq$1 to $\langle q_{ij}\rangle \simeq 0.13$ as $\epsilon$ decreases to 0. The static susceptibility, $\chi_{\text{stat}}(\epsilon)=N(\langle Q_{\text{stat}}^2\rangle-\langle Q_{\text{stat}}\rangle^2)$, has a peak at the value of $\epsilon$ where $\langle Q_{\text{stat}}\rangle$ drastically changes (the dashed vertical lines in Figure~\ref{phase} (A)). The width of the peaks  decreases dramatically  and the amplitudes increase as $T$ decreases, reflecting a sharp  change in $\langle Q_{\text{stat}}\rangle$ at low values of  $T$. Such dramatic changes in $\langle Q_{\text{stat}}\rangle$ and $\chi_{\text{stat}}(\epsilon)$ are particularly striking considering the relatively small size of the polymer, and provides evidence for the first-order like phase transition in the presence of the coupling field.

Next, we confirm that the abrupt change in $Q_{\text{stat}}$ indeed reflects a regular first order transition. In the first order transition, the minimum in the free energy $F_Q$ as a function of $Q_{\text{stat}}$ changes discontinuously at the phase transition point. The free energy, $F_Q$, is calculated  from the distribution $P(Q_{\text{stat}})$ as, 
$-\frac{NF_Q}{k_BT}=\langle \ln P(Q_{\text{stat}} )\rangle$.
In Figure~\ref{phase} (B), $F_Q$ at $T=0.35$ is plotted as a function of $Q_{\text{stat}}$ for various $\epsilon$. When $\epsilon=1$, $F_Q$ has minimum at $Q_{\text{stat}}\simeq0.93$, indicating that $\{\vec{r}\}$ stays around the metastable state associated with $\{\vec{r_0}\}$. As $\epsilon$ decreases to 0, the minimum shifts to $Q_{\text{stat}}=0.13$. At $\epsilon=0.574$, where $\chi_{\text{stat}}$ has a maximum, $F_Q$ has two minima at  $Q_{\text{stat}} \approx 0.34$ and $Q_{\text{stat}} \approx 0.84$.  This means that the two different states coexist at $\epsilon=0.574$, which leads to a discontinuous change in the order parameter. In Figure~\ref{phase} (A), we plot the minimum position of $F_Q$ as a function of $\epsilon$ for various $T$ (the solid lines). As shown in the graph, the minimum position changes discontinuously at $\epsilon$ where the fluctuation of $Q_{\text{stat}}$ is maximized, revealing the first order nature of the transition.

We arrive the same conclusion from the Ehrenfest classification of the phase transition. According to Ehrenfest classification, the first derivative of the free energy $F_\epsilon$ with respect to $\epsilon$ should be discontinuous at the transition point. We obtained $F_\epsilon$ from $F_Q$ by the Legendre transformation,  
\begin{eqnarray}
F_\epsilon = F_Q -\epsilon \ Q_{\text{stat}},
\label{legendre}
\end{eqnarray}  
where $\epsilon$ is equal to $\epsilon=\partial F_Q/\partial Q_{\text{stat}}$ \cite{FPbook}. Figure~\ref{phase} (D), which displays $F_\epsilon$ as a function of $\epsilon$ for various $T$, shows a discontinuous change in the slopes ($=dF_\epsilon/d\epsilon$) between -0.13 and -0.93 (the dashed and dashed-dotted lines, respectively), which is a signature of the first order transition.  Since $Q_{\text{stat}}=-\frac{dF_\epsilon}{d\epsilon}$ (Eq~(\ref{legendre})), we define the effective order parameters in the two states as $Q_{\text{liquid}}=0.13$ and $Q_{\text{glass}}=0.93$. Thus, the discontinuous change in the slope in figure~\ref{phase} (D) corresponds to the discontinuous change in $Q_{\text{stat}}$ between $Q_{\text{liquid}}$ and $Q_{\text{glass}}$. 
Figures~\ref{phase} (B) and (D) confirm that the first order transition in $Q_{\text{stat}}$ occurs with a change in $\epsilon$, thus verifying the existence of the metastable states in the SCP with $N=128$.  We also find that the first order transition  as $N$ increases, indicating that the metastable states should exist more stably in the SCP as $N\rightarrow\infty$ (section III in the SI ). 

RFOT theory predicts that metastable states, separated by barriers, should cease to exist above the dynamical transition temperature $T_d$, which implies that the first-order transition nature of $Q_{\text{stat}}$ should disappear at $T>T_d$. We found that $Q_{\text{stat}}$ changes continuously with $\epsilon$ when $T\geq1.8$ (see section III of the SI), indicating the disappearance of the metastable states at $T>T_d\simeq1.8$.   The collapse temperature is determined to be $T_\theta=2.3>T_d$ (section II in the SI). Thus, the equilibrium  collapse transition occurs before the dynamical transition, implying that the dynamics of the SCP in the temperature $T_d \le T \le T_\theta$ can be described by the standard polymer theory. Only below $T_d$ the dynamics is determined by activated transitions between equivalent compact structures.

The phase behavior  in Figure~\ref{phase} (A) is summarized  in Figure~\ref{phase} (C). The phase boundaries $(\epsilon_c,T_c)$ are associated with the peak positions in $\chi_{\text{stat}}$ at a given $T$ (the open symbols in the graph). At $T<T_d$ (the open symbols), $Q_{\text{stat}}$ changes discontinuously, and when $T>T_d$ (the filled triangles), $Q_{\text{stat}}$ changes continuously. Above $T_\theta$ (the black dashed lines), $\epsilon_c$ is less dependent on $T$. Because $T_c$ changes linearly with $\epsilon_c$ (when $T<T_d$), we extrapolate the phase boundary to $\epsilon=0$ (the blue solid line). The $y$ intercept in the linear fit indicates  $T_c$ at $\epsilon=0$. In the SCP, $T_c$ has a positive value at $\epsilon=0$ ($T_c(\epsilon=0)=0.14$), which signals that a thermodynamic transition (see below for the nature of the transition) would occur at the non-zero temperature when $\epsilon=0$. This means that the free energy of individual metastable states $F_\alpha$ is equal to the total free energy $F_{\text{tot}}$ at a finite temperature even without external fields, which is only possible if $S_{\text{conf}}=0$ at $T_c(\epsilon=0)$. 

We, therefore, investigated if $S_{\text{conf}}$  does indeed vanish at $T_c( \epsilon=0)$. Using two methods introduced elsewhere \cite{Berthier11668,Berth2019}, $S_{\text{conf}}$ at $\epsilon=0$ is estimated as a function of $T$.  Since the first order transition occurs when the entropic gain ($TS_{\text{conf}}$) is compensated by the energetic contribution of the fields ($N\epsilon_c (Q_{\text{glass}}-Q_{\text{liquid}}$)), we estimated the configurational entropy per monomer $s_{\text{conf}}=S_{\text{conf}}/N$ using  (Method 1) \cite{Berthier11668}, 
$Ts_{\text{conf}} = \epsilon_c [Q_{\text{glass}}-Q_{\text{liquid}}]$.
A more natural way is to calculate $TS_{\text{conf}}$ as the difference between $F_\alpha$ and $F_{\text{tot}}$, which corresponds to $F_{Q_{\text{glass}}}$ and $F_{Q_{\text{liquid}}}$ at $\epsilon=0$, respectively. Thus, $S_{\text{conf}}$ can also be calculated using,  
\begin{eqnarray}
Ts_{\text{conf}} = F_{Q_{\text{glass}}}(\epsilon=0)-F_{Q_{\text{liquid}}}(\epsilon=0).
\label{Method2}
\end{eqnarray}
Figure~\ref{phase} (E) plots $Ts_{\text{conf}}$ (the open symbols) as a function of $T$ using Method 1 and (Eq. \ref{Method2}). Linear extrapolation of $Ts_{\text{conf}}$ (the linear lines) shows that $s_{\text{conf}}$ vanishes at a similar non-zero value of $T$, regardless of which method is employed. It should be emphasized that the numerical value of the temperature is consistent with $T_c(\epsilon=0)$ (the open yellow star). This confirms that it is because the configurational entropy vanishes that the thermodynamic transition occurs at $T_c(\epsilon=0)$. 

What is the nature of the thermodynamic transition of the SCP at $T_c(\epsilon=0)$? First,  this corresponds to an ideal glass transition of the SCP. Most importantly, we note from Eq~(\ref{Method2}), $Ts_{\text{conf}}$ is the energy difference between the two states, which accounts for the latent heat at the transition. Therefore, Figure~\ref{phase} (E) shows that as $T$ approaches $T_c(\epsilon=0)$, the latent heat decreases and vanishes at $T_c(\epsilon=0)$. This implies that at  $T_c(\epsilon=0)$}, the SCP exhibits a {\it random} first order transition from liquid to an ideal glass without releasing latent heat but with a discontinuity in $Q_{\text{stat}}$. This is the fundamental feature of the ideal glass transition anticipated in the RFOT theory. Hence, Figures~\ref{phase} (C) and (E) show that the ideal glass transition in the SCP at $T_K=T_c(\epsilon=0) \ne 0$ is truly the analogue of $T_K$ in bulk glasses. 

Another major prediction made in our earlier studies is that there is a consistency between thermodynamic random first order transition and the underlying dynamics \cite{Kirkpatrick89JPhysA}. To investigate the dynamics of the SCP, we performed dynamic MC simulations (details in the SI). We define the time dependent overlap function $Q_{\text{dyn}}(t)$ as, 
\begin{eqnarray}
Q_{\text{dyn}} (t)=\frac{\sum _{(i,j)'}q_{ij}(t)q_{ij}(0)}{\sum_{(i,j)'}q_{ij}(0)},
\label{contact}
\end{eqnarray}
where $q_{ij}(t)$ is the contact function of a single polymer at time $t$; $Q_{\text{dyn}}(t)$ quantifies how rapidly the contact map loses memory of the initial pattern. By definition $Q_{\text{dyn}}(t=0)=1$.  As $t\rightarrow\infty$, the SCP loses memory of the structural correlation, and thus the pattern of the contact map also becomes independent of the initial state. Consequently, $Q_{\text{dyn}}(t)$ decays to $\langle q_{ij}\rangle$.

\begin {figure}
\centering
\includegraphics  [width=3.3in] {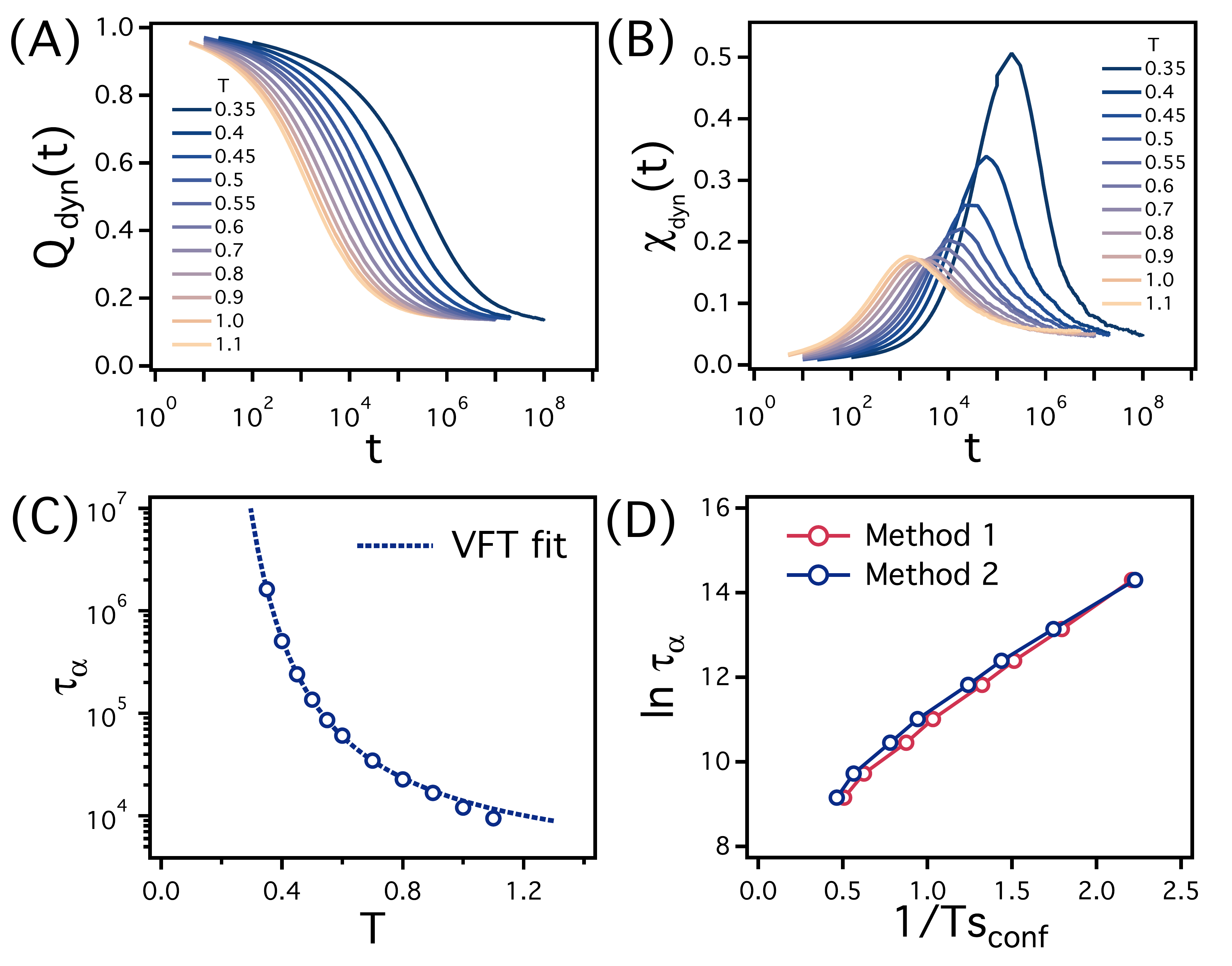}
\caption{\textbf{Glassy dynamics in the SCP.} (A) Time dependent overlap function $Q_{\text{dyn}}(t)$ and (B) susceptibility $\chi_{\text{dyn}}(t)$ for various $T$. (C) Dependence of $\tau_\alpha$ on $T$ (the open symbols). The dashed line is the VFT fit. (D) Relation between $\tau_\alpha$ and $s_{\text{conf}}$. }
\label{dynamics}
\end{figure}

Figure~\ref{dynamics} (A) shows $Q_{\text{dyn}}(t)$ as a function of $t$ at different $T$. As $T$ decreases from 1.1 to 0.35, $Q_{\text{dyn}}(t)$ decays more slowly with the decay time constant increasing by a few orders of magnitude.  Figure~\ref{dynamics} (B) shows that heights and timescales of the peak in the dynamic susceptibility, $\chi_{\text{dyn}}$, $\chi_{\text{dyn}}(t) = N[\langle Q_{\text{dyn}} (t)^2\rangle-\langle Q_{\text{dyn}} (t)\rangle^2]$ increase as $T$ decreases, implying that the structural relaxation becomes heterogeneous as $T$ decreases \cite{PhysRevA.37.4439,Flenner:2011ht,Flenner:2010io}. Thus, Figures~\ref{dynamics} (A) and (B) reveal that the sluggish structural relaxation in the SCP is accompanied by enhanced dynamic heterogeneity, which is an important dynamic property of glassy liquids \cite{EdigerMD1996,EdigerMD2000,BiroliG2013}. 

We estimated the structural relaxation time $\tau_\alpha$ using $Q_{\text{dyn}}(t=\tau_\alpha)=0.3$. Figure~\ref{dynamics} (C) shows that $\tau_\alpha$ increases by more than two orders of magnitude when $T$ decreases from 1.1 to 0.35 (the open symbols). Such a large increase in $\tau_\alpha$ in a finite sized polymer  is substantial. The sharp increase in $\tau_\alpha$ in the SCP as well as in metallic \cite{BuschR1998}, colloidal systems \cite{cho2019fragiletostrong}, and molecular glasses \cite{Angell1924}, is described well by the Vogel-Tamman-Flucher (VFT) equation, $\tau_\alpha=\tau_0\exp[\frac{D_0T_0}{T-T_0}]$, where $\tau_0$, $D_0$ and $T_0$ are fitting parameters. We fit the calculated $\tau_\alpha$ as a function of $T$ to the VFT equation (the dashed line in Figure~\ref{dynamics} (C)), which gives $T_0=0.1$, (the green open symbols in Figure~\ref{phase} (C) and (E) the temperature at  which $\tau_\alpha$ diverges. The value of $T_0$ is close to $T_K$. 

The divergence of $\tau_\alpha$ is closely associated with the decrease in $S_{\text{conf}}$.  A scaling relation exists between $\tau_\alpha$ and $S_{\text{conf}}$ of bulk liquids, $\ln \tau_\alpha\sim1/Ts_{\text{conf}}$ \cite{PhysRevA.40.1045}. In Figure~\ref{dynamics} (D), we plot $\ln \tau_\alpha$ as a function of $Ts_{\text{conf}}$ obtained using Method 1 and (\ref{Method2}). There is a modest deviation from a linear line  possibly due to the confined geometry of the SCP. Nevertheless, this graph clearly shows that the increase in $\tau_\alpha$ is closely related to the decrease in $s_{\text{conf}}$, implying a causal relationship between the two quantities in the finite-sized SCP. Thus, the dynamics and statics of the SCP are consistently  predicted by the RFOT theory.

The glassy behavior illustrated  in Figures~\ref{phase} and \ref{dynamics} for the SCP is shared by bulk liquid glasses. Amorphous metastable states in bulk liquid glasses was also probed by the FP method \cite{PhysRevE.88.022313,PhysRevE.89.022309,PhysRevLett.114.205701,Berthier11668}.  And, the same relation between configurational entropy and relaxation time was found. The fluctuations of order parameters associated with the SCP appears to belong to the  random field Ising model (RFIM) universality class (section IV in the SI) in the presence of the quenched field. 

Our findings show that the essence of the RFOT theory holds for a diverse systems that undergo liquid to glass transition. It appears that the difficulty in accessing the ideal glass transition temperature in atomic glasses is greatly lessened in the SCP. The use of the coupling field allows us to probe the energy landscape readily. We will be remiss if we do not stress that the random first order transition at $T_K$, with $\epsilon =0$ that we establish by suitable numerical methods within the FP scheme, differs from the line of first order transitions that occur at $\epsilon_c \ne 0$.  In the SCP, the transition is from a disordered state to an ordered compact state. The dynamics in the disordered state can be described using well-known theory above $T_d$ whereas below $T_d$ polymer dynamics requires RFOT theory for capturing the activated transition. Finally, the SCP, and more importantly the much larger chromosomes that are compact and exhibit glass-like dynamics \cite{Kang05PRL}, could be ideal systems to explore RFOT concepts. 
 

{\bf Acknowledgements:} This work was supported by NSF (CHE 19-00093), and the Welch Foundation Grant F-0019 through the Collie-Welch chair.



\begin{thebibliography}{13}%
\makeatletter
\providecommand \@ifxundefined [1]{%
 \@ifx{#1\undefined}
}%
\providecommand \@ifnum [1]{%
 \ifnum #1\expandafter \@firstoftwo
 \else \expandafter \@secondoftwo
 \fi
}%
\providecommand \@ifx [1]{%
 \ifx #1\expandafter \@firstoftwo
 \else \expandafter \@secondoftwo
 \fi
}%
\providecommand \natexlab [1]{#1}%
\providecommand \enquote  [1]{``#1''}%
\providecommand \bibnamefont  [1]{#1}%
\providecommand \bibfnamefont [1]{#1}%
\providecommand \citenamefont [1]{#1}%
\providecommand \href@noop [0]{\@secondoftwo}%
\providecommand \href [0]{\begingroup \@sanitize@url \@href}%
\providecommand \@href[1]{\@@startlink{#1}\@@href}%
\providecommand \@@href[1]{\endgroup#1\@@endlink}%
\providecommand \@sanitize@url [0]{\catcode `\\12\catcode `\$12\catcode
  `\&12\catcode `\#12\catcode `\^12\catcode `\_12\catcode `\%12\relax}%
\providecommand \@@startlink[1]{}%
\providecommand \@@endlink[0]{}%
\providecommand \url  [0]{\begingroup\@sanitize@url \@url }%
\providecommand \@url [1]{\endgroup\@href {#1}{\urlprefix }}%
\providecommand \urlprefix  [0]{URL }%
\providecommand \Eprint [0]{\href }%
\providecommand \doibase [0]{http://dx.doi.org/}%
\providecommand \selectlanguage [0]{\@gobble}%
\providecommand \bibinfo  [0]{\@secondoftwo}%
\providecommand \bibfield  [0]{\@secondoftwo}%
\providecommand \translation [1]{[#1]}%
\providecommand \BibitemOpen [0]{}%
\providecommand \bibitemStop [0]{}%
\providecommand \bibitemNoStop [0]{.\EOS\space}%
\providecommand \EOS [0]{\spacefactor3000\relax}%
\providecommand \BibitemShut  [1]{\csname bibitem#1\endcsname}%
\let\auto@bib@innerbib\@empty
\bibitem [{\citenamefont {Frenkel}\ and\ \citenamefont
  {Smit}(1996)}]{FrenkelSmit}%
  \BibitemOpen
  \bibfield  {author} {\bibinfo {author} {\bibfnamefont {D.}~\bibnamefont
  {Frenkel}}\ and\ \bibinfo {author} {\bibfnamefont {B.}~\bibnamefont {Smit}},\
  }\href@noop {} {\emph {\bibinfo {title} {Understanding molecular simulation:
  from algorithms to applications.}}},\ \bibinfo {edition} {2nd}\ ed.\
  (\bibinfo  {publisher} {Academic Press},\ \bibinfo {year} {1996})\BibitemShut
  {NoStop}%
\bibitem [{\citenamefont {Berthier}\ \emph {et~al.}(2017)\citenamefont
  {Berthier}, \citenamefont {Charbonneau}, \citenamefont {Coslovich},
  \citenamefont {Ninarello}, \citenamefont {Ozawa},\ and\ \citenamefont
  {Yaida}}]{Berthier11356}%
  \BibitemOpen
  \bibfield  {author} {\bibinfo {author} {\bibfnamefont {L.}~\bibnamefont
  {Berthier}}, \bibinfo {author} {\bibfnamefont {P.}~\bibnamefont
  {Charbonneau}}, \bibinfo {author} {\bibfnamefont {D.}~\bibnamefont
  {Coslovich}}, \bibinfo {author} {\bibfnamefont {A.}~\bibnamefont
  {Ninarello}}, \bibinfo {author} {\bibfnamefont {M.}~\bibnamefont {Ozawa}}, \
  and\ \bibinfo {author} {\bibfnamefont {S.}~\bibnamefont {Yaida}},\ }\bibfield
   {title} {\enquote {\bibinfo {title} {Configurational entropy measurements in
  extremely supercooled liquids that break the glass ceiling},}\ }\href@noop {}
  {\bibfield  {journal} {\bibinfo  {journal} {Proc. Natl. Acad. Sci. USA}\
  }\textbf {\bibinfo {volume} {114}},\ \bibinfo {pages} {11356--11361}
  (\bibinfo {year} {2017})}\BibitemShut {NoStop}%
\bibitem [{\citenamefont {Berthier}(2013)}]{PhysRevE.88.022313}%
  \BibitemOpen
  \bibfield  {author} {\bibinfo {author} {\bibfnamefont {L.}~\bibnamefont
  {Berthier}},\ }\bibfield  {title} {\enquote {\bibinfo {title} {Overlap
  fluctuations in glass-forming liquids},}\ }\href@noop {} {\bibfield
  {journal} {\bibinfo  {journal} {Phys. Rev. E}\ }\textbf {\bibinfo {volume}
  {88}},\ \bibinfo {pages} {022313} (\bibinfo {year} {2013})}\BibitemShut
  {NoStop}%
\bibitem [{\citenamefont {Berthier}\ and\ \citenamefont
  {Coslovich}(2014)}]{Berthier11668}%
  \BibitemOpen
  \bibfield  {author} {\bibinfo {author} {\bibfnamefont {L.}~\bibnamefont
  {Berthier}}\ and\ \bibinfo {author} {\bibfnamefont {D.}~\bibnamefont
  {Coslovich}},\ }\bibfield  {title} {\enquote {\bibinfo {title} {Novel
  approach to numerical measurements of the configurational entropy in
  supercooled liquids},}\ }\href@noop {} {\bibfield  {journal} {\bibinfo
  {journal} {Proc. Natl. Acad. Sci. USA}\ }\textbf {\bibinfo {volume} {111}},\
  \bibinfo {pages} {11668--11672} (\bibinfo {year} {2014})}\BibitemShut
  {NoStop}%
\bibitem [{\citenamefont {Taylor}\ \emph {et~al.}(2009)\citenamefont {Taylor},
  \citenamefont {Paul},\ and\ \citenamefont {Binder}}]{jcpsingle1}%
  \BibitemOpen
  \bibfield  {author} {\bibinfo {author} {\bibfnamefont {M.~P.}\ \bibnamefont
  {Taylor}}, \bibinfo {author} {\bibfnamefont {W.}~\bibnamefont {Paul}}, \ and\
  \bibinfo {author} {\bibfnamefont {K.}~\bibnamefont {Binder}},\ }\bibfield
  {title} {\enquote {\bibinfo {title} {Phase transitions of a single polymer
  chain: A wang-landau simulation study},}\ }\href@noop {} {\bibfield
  {journal} {\bibinfo  {journal} {J. Chem. Phys.}\ }\textbf {\bibinfo {volume}
  {131}},\ \bibinfo {pages} {114907} (\bibinfo {year} {2009})}\BibitemShut
  {NoStop}%
\bibitem [{\citenamefont {Parsons}\ and\ \citenamefont
  {Williams}(2006)}]{jcpsingle2}%
  \BibitemOpen
  \bibfield  {author} {\bibinfo {author} {\bibfnamefont {D.~F.}\ \bibnamefont
  {Parsons}}\ and\ \bibinfo {author} {\bibfnamefont {D.~R.~M.}\ \bibnamefont
  {Williams}},\ }\bibfield  {title} {\enquote {\bibinfo {title} {An off-lattice
  wang-landau study of the coil-globule and melting transitions of a flexible
  homopolymer},}\ }\href@noop {} {\bibfield  {journal} {\bibinfo  {journal}
  {The Journal of Chemical Physics}\ }\textbf {\bibinfo {volume} {124}},\
  \bibinfo {pages} {221103} (\bibinfo {year} {2006})}\BibitemShut {NoStop}%
\bibitem [{\citenamefont {Kirkpatrick}\ \emph {et~al.}(1989)\citenamefont
  {Kirkpatrick}, \citenamefont {Thirumalai},\ and\ \citenamefont
  {Wolynes}}]{PhysRevA.40.1045}%
  \BibitemOpen
  \bibfield  {author} {\bibinfo {author} {\bibfnamefont {T.~R.}\ \bibnamefont
  {Kirkpatrick}}, \bibinfo {author} {\bibfnamefont {D.}~\bibnamefont
  {Thirumalai}}, \ and\ \bibinfo {author} {\bibfnamefont {P.~G.}\ \bibnamefont
  {Wolynes}},\ }\bibfield  {title} {\enquote {\bibinfo {title} {Scaling
  concepts for the dynamics of viscous liquids near an ideal glassy state},}\
  }\href@noop {} {\bibfield  {journal} {\bibinfo  {journal} {Phys. Rev. A}\
  }\textbf {\bibinfo {volume} {40}},\ \bibinfo {pages} {1045--1054} (\bibinfo
  {year} {1989})}\BibitemShut {NoStop}%
\bibitem [{\citenamefont {Franz}\ and\ \citenamefont
  {Parisi}(1997)}]{PhysRevLett.79.2486}%
  \BibitemOpen
  \bibfield  {author} {\bibinfo {author} {\bibfnamefont {S.}~\bibnamefont
  {Franz}}\ and\ \bibinfo {author} {\bibfnamefont {G.}~\bibnamefont {Parisi}},\
  }\bibfield  {title} {\enquote {\bibinfo {title} {Phase diagram of coupled
  glassy systems: a mean-field study},}\ }\href@noop {} {\bibfield  {journal}
  {\bibinfo  {journal} {Phys. Rev. Lett.}\ }\textbf {\bibinfo {volume} {79}},\
  \bibinfo {pages} {2486--2489} (\bibinfo {year} {1997})}\BibitemShut {NoStop}%
\bibitem [{\citenamefont {Berthier}\ and\ \citenamefont
  {Jack}(2015)}]{PhysRevLett.114.205701}%
  \BibitemOpen
  \bibfield  {author} {\bibinfo {author} {\bibfnamefont {L.}~\bibnamefont
  {Berthier}}\ and\ \bibinfo {author} {\bibfnamefont {R.~L.}\ \bibnamefont
  {Jack}},\ }\bibfield  {title} {\enquote {\bibinfo {title} {Evidence for a
  disordered critical point in a glass-forming liquid},}\ }\href@noop {}
  {\bibfield  {journal} {\bibinfo  {journal} {Phys. Rev. Lett.}\ }\textbf
  {\bibinfo {volume} {114}},\ \bibinfo {pages} {205701} (\bibinfo {year}
  {2015})}\BibitemShut {NoStop}%
\bibitem [{\citenamefont {Berthier}\ \emph {et~al.}(2005)\citenamefont
  {Berthier}, \citenamefont {Biroli}, \citenamefont {Bouchaud}, \citenamefont
  {Cipelletti}, \citenamefont {Masri}, \citenamefont
  {L{\textquoteright}H{\^o}te}, \citenamefont {Ladieu},\ and\ \citenamefont
  {Pierno}}]{Berthier1797}%
  \BibitemOpen
  \bibfield  {author} {\bibinfo {author} {\bibfnamefont {L.}~\bibnamefont
  {Berthier}}, \bibinfo {author} {\bibfnamefont {G.}~\bibnamefont {Biroli}},
  \bibinfo {author} {\bibfnamefont {J.-P.}\ \bibnamefont {Bouchaud}}, \bibinfo
  {author} {\bibfnamefont {L.}~\bibnamefont {Cipelletti}}, \bibinfo {author}
  {\bibfnamefont {D.~El}\ \bibnamefont {Masri}}, \bibinfo {author}
  {\bibfnamefont {D.}~\bibnamefont {L{\textquoteright}H{\^o}te}}, \bibinfo
  {author} {\bibfnamefont {F.}~\bibnamefont {Ladieu}}, \ and\ \bibinfo {author}
  {\bibfnamefont {M.}~\bibnamefont {Pierno}},\ }\bibfield  {title} {\enquote
  {\bibinfo {title} {Direct experimental evidence of a growing length scale
  accompanying the glass transition},}\ }\href@noop {} {\bibfield  {journal}
  {\bibinfo  {journal} {Science}\ }\textbf {\bibinfo {volume} {310}},\ \bibinfo
  {pages} {1797--1800} (\bibinfo {year} {2005})}\BibitemShut {NoStop}%
\bibitem [{\citenamefont {Berthier}\ \emph
  {et~al.}(2007{\natexlab{a}})\citenamefont {Berthier}, \citenamefont {Biroli},
  \citenamefont {Bouchaud}, \citenamefont {Kob}, \citenamefont {Miyazaki},\
  and\ \citenamefont {Reichman}}]{berthiercritical1}%
  \BibitemOpen
  \bibfield  {author} {\bibinfo {author} {\bibfnamefont {L.}~\bibnamefont
  {Berthier}}, \bibinfo {author} {\bibfnamefont {G.}~\bibnamefont {Biroli}},
  \bibinfo {author} {\bibfnamefont {J.-P.}\ \bibnamefont {Bouchaud}}, \bibinfo
  {author} {\bibfnamefont {W.}~\bibnamefont {Kob}}, \bibinfo {author}
  {\bibfnamefont {K.}~\bibnamefont {Miyazaki}}, \ and\ \bibinfo {author}
  {\bibfnamefont {D.~R.}\ \bibnamefont {Reichman}},\ }\bibfield  {title}
  {\enquote {\bibinfo {title} {Spontaneous and induced dynamic fluctuations in
  glass formers. i. general results and dependence on ensemble and dynamics},}\
  }\href@noop {} {\bibfield  {journal} {\bibinfo  {journal} {J. Chem. Phys.}\
  }\textbf {\bibinfo {volume} {126}},\ \bibinfo {pages} {184503} (\bibinfo
  {year} {2007}{\natexlab{a}})}\BibitemShut {NoStop}%
\bibitem [{\citenamefont {Berthier}\ \emph
  {et~al.}(2007{\natexlab{b}})\citenamefont {Berthier}, \citenamefont {Biroli},
  \citenamefont {Bouchaud}, \citenamefont {Kob}, \citenamefont {Miyazaki},\
  and\ \citenamefont {Reichman}}]{berthiercritical2}%
  \BibitemOpen
  \bibfield  {author} {\bibinfo {author} {\bibfnamefont {L.}~\bibnamefont
  {Berthier}}, \bibinfo {author} {\bibfnamefont {G.}~\bibnamefont {Biroli}},
  \bibinfo {author} {\bibfnamefont {J.-P.}\ \bibnamefont {Bouchaud}}, \bibinfo
  {author} {\bibfnamefont {W.}~\bibnamefont {Kob}}, \bibinfo {author}
  {\bibfnamefont {K.}~\bibnamefont {Miyazaki}}, \ and\ \bibinfo {author}
  {\bibfnamefont {D.~R.}\ \bibnamefont {Reichman}},\ }\bibfield  {title}
  {\enquote {\bibinfo {title} {Spontaneous and induced dynamic correlations in
  glass formers. ii. model calculations and comparison to numerical
  simulations},}\ }\href@noop {} {\bibfield  {journal} {\bibinfo  {journal}
  {The Journal of Chemical Physics}\ }\textbf {\bibinfo {volume} {126}},\
  \bibinfo {pages} {184504} (\bibinfo {year} {2007}{\natexlab{b}})}\BibitemShut
  {NoStop}%
\bibitem [{\citenamefont {Fisher}()}]{PhysRevLett.56.416}%
  \BibitemOpen
  \bibfield  {author} {\bibinfo {author} {\bibfnamefont {D.~S.}\ \bibnamefont
  {Fisher}},\ }\bibfield  {title} {\enquote {\bibinfo {title} {Scaling and
  critical slowing down in random-field ising systems},}\ }\href@noop {}
  {\bibfield  {journal} {\bibinfo  {journal} {Phys. Rev. Lett.}\ }\textbf
  {\bibinfo {volume} {56}},\ \bibinfo {pages} {416--419}}\BibitemShut {NoStop}%
\end{thebibliography}%


\begin{thebibliography}{43}
\expandafter\ifx\csname natexlab\endcsname\relax\def\natexlab#1{#1}\fi
\expandafter\ifx\csname bibnamefont\endcsname\relax
  \def\bibnamefont#1{#1}\fi
\expandafter\ifx\csname bibfnamefont\endcsname\relax
  \def\bibfnamefont#1{#1}\fi
\expandafter\ifx\csname citenamefont\endcsname\relax
  \def\citenamefont#1{#1}\fi
\expandafter\ifx\csname url\endcsname\relax
  \def\url#1{\texttt{#1}}\fi
\expandafter\ifx\csname urlprefix\endcsname\relax\def\urlprefix{URL }\fi
\providecommand{\bibinfo}[2]{#2}
\providecommand{\eprint}[2][]{\url{#2}}

\bibitem[{\citenamefont{Finn and Misteli}(2019)}]{Finn19Science}
\bibinfo{author}{\bibfnamefont{E.~H.} \bibnamefont{Finn}} \bibnamefont{and}
  \bibinfo{author}{\bibfnamefont{T.}~\bibnamefont{Misteli}},
  \bibinfo{journal}{Science} \textbf{\bibinfo{volume}{365}}
  (\bibinfo{year}{2019}).

\bibitem[{\citenamefont{Bronshtein et~al.}({2015})\citenamefont{Bronshtein,
  Kepten, Kanter, Berezin, Lindner, Redwood, Mai, Gonzalo, Foisner, Shav-Tal
  et~al.}}]{Bronshtein15NatComm}
\bibinfo{author}{\bibfnamefont{I.}~\bibnamefont{Bronshtein}},
  \bibinfo{author}{\bibfnamefont{E.}~\bibnamefont{Kepten}},
  \bibinfo{author}{\bibfnamefont{I.}~\bibnamefont{Kanter}},
  \bibinfo{author}{\bibfnamefont{S.}~\bibnamefont{Berezin}},
  \bibinfo{author}{\bibfnamefont{M.}~\bibnamefont{Lindner}},
  \bibinfo{author}{\bibfnamefont{A.~B.} \bibnamefont{Redwood}},
  \bibinfo{author}{\bibfnamefont{S.}~\bibnamefont{Mai}},
  \bibinfo{author}{\bibfnamefont{S.}~\bibnamefont{Gonzalo}},
  \bibinfo{author}{\bibfnamefont{R.}~\bibnamefont{Foisner}},
  \bibinfo{author}{\bibfnamefont{Y.}~\bibnamefont{Shav-Tal}},
  \bibnamefont{et~al.}, \bibinfo{journal}{{Nat. Comm.}}
  \textbf{\bibinfo{volume}{{6}}} (\bibinfo{year}{{2015}}).

\bibitem[{\citenamefont{Berndsen et~al.}(2014)\citenamefont{Berndsen, Keller,
  Grimes, Jardine, and Smith}}]{Berndsen8345}
\bibinfo{author}{\bibfnamefont{Z.~T.} \bibnamefont{Berndsen}},
  \bibinfo{author}{\bibfnamefont{N.}~\bibnamefont{Keller}},
  \bibinfo{author}{\bibfnamefont{S.}~\bibnamefont{Grimes}},
  \bibinfo{author}{\bibfnamefont{P.~J.} \bibnamefont{Jardine}},
  \bibnamefont{and} \bibinfo{author}{\bibfnamefont{D.~E.} \bibnamefont{Smith}},
  \bibinfo{journal}{Proc. Natl. Acad. Sci. USA} \textbf{\bibinfo{volume}{111}},
  \bibinfo{pages}{8345} (\bibinfo{year}{2014}).

\bibitem[{\citenamefont{Keller et~al.}(2016)\citenamefont{Keller, Grimes,
  Jardine, and Smith}}]{Keller:2016bh}
\bibinfo{author}{\bibfnamefont{N.}~\bibnamefont{Keller}},
  \bibinfo{author}{\bibfnamefont{S.}~\bibnamefont{Grimes}},
  \bibinfo{author}{\bibfnamefont{P.~J.} \bibnamefont{Jardine}},
  \bibnamefont{and} \bibinfo{author}{\bibfnamefont{D.~E.} \bibnamefont{Smith}},
  \bibinfo{journal}{Nat. Phys.} \textbf{\bibinfo{volume}{12}},
  \bibinfo{pages}{757} (\bibinfo{year}{2016}).

\bibitem[{\citenamefont{Kwon et~al.}(2017)\citenamefont{Kwon, Cho, Kim, and
  Sung}}]{PhysRevLett.119.087801}
\bibinfo{author}{\bibfnamefont{S.}~\bibnamefont{Kwon}},
  \bibinfo{author}{\bibfnamefont{H.~W.} \bibnamefont{Cho}},
  \bibinfo{author}{\bibfnamefont{J.}~\bibnamefont{Kim}}, \bibnamefont{and}
  \bibinfo{author}{\bibfnamefont{B.~J.} \bibnamefont{Sung}},
  \bibinfo{journal}{Phys. Rev. Lett.} \textbf{\bibinfo{volume}{119}},
  \bibinfo{pages}{087801} (\bibinfo{year}{2017}).

\bibitem[{\citenamefont{Morgan et~al.}({2020})\citenamefont{Morgan, Avinery,
  Rahamim, Beck, and Saleh}}]{Morgan20PRL}
\bibinfo{author}{\bibfnamefont{I.~L.} \bibnamefont{Morgan}},
  \bibinfo{author}{\bibfnamefont{R.}~\bibnamefont{Avinery}},
  \bibinfo{author}{\bibfnamefont{G.}~\bibnamefont{Rahamim}},
  \bibinfo{author}{\bibfnamefont{R.}~\bibnamefont{Beck}}, \bibnamefont{and}
  \bibinfo{author}{\bibfnamefont{O.~A.} \bibnamefont{Saleh}},
  \bibinfo{journal}{{Phys. Rev. Lett.}} \textbf{\bibinfo{volume}{{125}}},
  \bibinfo{pages}{{058001}} (\bibinfo{year}{{2020}}).

\bibitem[{\citenamefont{Perkins et~al.}(1994)\citenamefont{Perkins, Quake,
  Smith, and Chu}}]{Perkins822}
\bibinfo{author}{\bibfnamefont{T.~T.} \bibnamefont{Perkins}},
  \bibinfo{author}{\bibfnamefont{S.~R.} \bibnamefont{Quake}},
  \bibinfo{author}{\bibfnamefont{D.~E.} \bibnamefont{Smith}}, \bibnamefont{and}
  \bibinfo{author}{\bibfnamefont{S.}~\bibnamefont{Chu}},
  \bibinfo{journal}{Science} \textbf{\bibinfo{volume}{264}},
  \bibinfo{pages}{822} (\bibinfo{year}{1994}).

\bibitem[{\citenamefont{Latinwo and Schroeder}(2011)}]{C1SM05298E}
\bibinfo{author}{\bibfnamefont{F.}~\bibnamefont{Latinwo}} \bibnamefont{and}
  \bibinfo{author}{\bibfnamefont{C.~M.} \bibnamefont{Schroeder}},
  \bibinfo{journal}{Soft Matter} \textbf{\bibinfo{volume}{7}},
  \bibinfo{pages}{7907} (\bibinfo{year}{2011}).

\bibitem[{\citenamefont{Tamm et~al.}(2015)\citenamefont{Tamm, Nazarov,
  Gavrilov, and Chertovich}}]{PhysRevLett.114.178102}
\bibinfo{author}{\bibfnamefont{M.~V.} \bibnamefont{Tamm}},
  \bibinfo{author}{\bibfnamefont{L.~I.} \bibnamefont{Nazarov}},
  \bibinfo{author}{\bibfnamefont{A.~A.} \bibnamefont{Gavrilov}},
  \bibnamefont{and} \bibinfo{author}{\bibfnamefont{A.~V.}
  \bibnamefont{Chertovich}}, \bibinfo{journal}{Phys. Rev. Lett.}
  \textbf{\bibinfo{volume}{114}}, \bibinfo{pages}{178102}
  (\bibinfo{year}{2015}).

\bibitem[{\citenamefont{Yoshikawa et~al.}(1996)\citenamefont{Yoshikawa,
  Takahashi, Vasilevskaya, and Khokhlov}}]{PhysRevLett.76.3029}
\bibinfo{author}{\bibfnamefont{K.}~\bibnamefont{Yoshikawa}},
  \bibinfo{author}{\bibfnamefont{M.}~\bibnamefont{Takahashi}},
  \bibinfo{author}{\bibfnamefont{V.~V.} \bibnamefont{Vasilevskaya}},
  \bibnamefont{and} \bibinfo{author}{\bibfnamefont{A.~R.}
  \bibnamefont{Khokhlov}}, \bibinfo{journal}{Phys. Rev. Lett.}
  \textbf{\bibinfo{volume}{76}}, \bibinfo{pages}{3029} (\bibinfo{year}{1996}).

\bibitem[{\citenamefont{Rampf et~al.}(2006)\citenamefont{Rampf, Binder, and
  Paul}}]{Rampf2006}
\bibinfo{author}{\bibfnamefont{F.}~\bibnamefont{Rampf}},
  \bibinfo{author}{\bibfnamefont{K.}~\bibnamefont{Binder}}, \bibnamefont{and}
  \bibinfo{author}{\bibfnamefont{W.}~\bibnamefont{Paul}}, \bibinfo{journal}{J.
  Polym. Sci. B Polym. Phys.} \textbf{\bibinfo{volume}{44}},
  \bibinfo{pages}{2542} (\bibinfo{year}{2006}).

\bibitem[{\citenamefont{Taylor et~al.}(2009)\citenamefont{Taylor, Paul, and
  Binder}}]{Taylor2009}
\bibinfo{author}{\bibfnamefont{M.~P.} \bibnamefont{Taylor}},
  \bibinfo{author}{\bibfnamefont{W.}~\bibnamefont{Paul}}, \bibnamefont{and}
  \bibinfo{author}{\bibfnamefont{K.}~\bibnamefont{Binder}},
  \bibinfo{journal}{J. Chem. Phys.} \textbf{\bibinfo{volume}{131}},
  \bibinfo{pages}{114907} (\bibinfo{year}{2009}).

\bibitem[{\citenamefont{Dokholyan et~al.}(2002)\citenamefont{Dokholyan, Pitard,
  Buldyrev, and Stanley}}]{PhysRevE.65.030801}
\bibinfo{author}{\bibfnamefont{N.~V.} \bibnamefont{Dokholyan}},
  \bibinfo{author}{\bibfnamefont{E.}~\bibnamefont{Pitard}},
  \bibinfo{author}{\bibfnamefont{S.~V.} \bibnamefont{Buldyrev}},
  \bibnamefont{and} \bibinfo{author}{\bibfnamefont{H.~E.}
  \bibnamefont{Stanley}}, \bibinfo{journal}{Phys. Rev. E}
  \textbf{\bibinfo{volume}{65}}, \bibinfo{pages}{030801}
  (\bibinfo{year}{2002}).

\bibitem[{\citenamefont{Tress et~al.}(2013)\citenamefont{Tress, Mapesa,
  Kossack, Kipnusu, Reiche, and Kremer}}]{Tress1371}
\bibinfo{author}{\bibfnamefont{M.}~\bibnamefont{Tress}},
  \bibinfo{author}{\bibfnamefont{E.~U.} \bibnamefont{Mapesa}},
  \bibinfo{author}{\bibfnamefont{W.}~\bibnamefont{Kossack}},
  \bibinfo{author}{\bibfnamefont{W.~K.} \bibnamefont{Kipnusu}},
  \bibinfo{author}{\bibfnamefont{M.}~\bibnamefont{Reiche}}, \bibnamefont{and}
  \bibinfo{author}{\bibfnamefont{F.}~\bibnamefont{Kremer}},
  \bibinfo{journal}{Science} \textbf{\bibinfo{volume}{341}},
  \bibinfo{pages}{1371} (\bibinfo{year}{2013}).

\bibitem[{\citenamefont{Merling et~al.}(2016)\citenamefont{Merling, Mileski,
  Douglas, and Simmons}}]{macr6b01461}
\bibinfo{author}{\bibfnamefont{W.~L.} \bibnamefont{Merling}},
  \bibinfo{author}{\bibfnamefont{J.~B.} \bibnamefont{Mileski}},
  \bibinfo{author}{\bibfnamefont{J.~F.} \bibnamefont{Douglas}},
  \bibnamefont{and} \bibinfo{author}{\bibfnamefont{D.~S.}
  \bibnamefont{Simmons}}, \bibinfo{journal}{Macromolecules}
  \textbf{\bibinfo{volume}{49}}, \bibinfo{pages}{7597} (\bibinfo{year}{2016}).

\bibitem[{\citenamefont{Orland et~al.}(1985)\citenamefont{Orland, Itzykson, and
  Dedominicis}}]{Orland85JPhys}
\bibinfo{author}{\bibfnamefont{H.}~\bibnamefont{Orland}},
  \bibinfo{author}{\bibfnamefont{C.}~\bibnamefont{Itzykson}}, \bibnamefont{and}
  \bibinfo{author}{\bibfnamefont{C.}~\bibnamefont{Dedominicis}},
  \bibinfo{journal}{J. de Physique Lett.} \textbf{\bibinfo{volume}{46}},
  \bibinfo{pages}{L353} (\bibinfo{year}{1985}).

\bibitem[{\citenamefont{Kirkpatrick et~al.}(1989)\citenamefont{Kirkpatrick,
  Thirumalai, and Wolynes}}]{PhysRevA.40.1045}
\bibinfo{author}{\bibfnamefont{T.~R.} \bibnamefont{Kirkpatrick}},
  \bibinfo{author}{\bibfnamefont{D.}~\bibnamefont{Thirumalai}},
  \bibnamefont{and} \bibinfo{author}{\bibfnamefont{P.~G.}
  \bibnamefont{Wolynes}}, \bibinfo{journal}{Phys. Rev. A}
  \textbf{\bibinfo{volume}{40}}, \bibinfo{pages}{1045} (\bibinfo{year}{1989}).

\bibitem[{\citenamefont{Kirkpatrick and
  Thirumalai}(1987{\natexlab{a}})}]{Kirkpatrick87PRL}
\bibinfo{author}{\bibfnamefont{T.~R.} \bibnamefont{Kirkpatrick}}
  \bibnamefont{and}
  \bibinfo{author}{\bibfnamefont{D.}~\bibnamefont{Thirumalai}},
  \bibinfo{journal}{Phys. Rev. Lett.} \textbf{\bibinfo{volume}{58}},
  \bibinfo{pages}{2091} (\bibinfo{year}{1987}{\natexlab{a}}).

\bibitem[{\citenamefont{Kirkpatrick and
  Thirumalai}(1987{\natexlab{b}})}]{Kirkpatrick87PRB}
\bibinfo{author}{\bibfnamefont{T.~R.} \bibnamefont{Kirkpatrick}}
  \bibnamefont{and}
  \bibinfo{author}{\bibfnamefont{D.}~\bibnamefont{Thirumalai}},
  \bibinfo{journal}{Phys. Rev. B} \textbf{\bibinfo{volume}{36}},
  \bibinfo{pages}{5388} (\bibinfo{year}{1987}{\natexlab{b}}).

\bibitem[{\citenamefont{Kirkpatrick and
  Wolynes}(1987)}]{Kirkpatrick-Wolynes87PRB}
\bibinfo{author}{\bibfnamefont{T.~R.} \bibnamefont{Kirkpatrick}}
  \bibnamefont{and} \bibinfo{author}{\bibfnamefont{P.~G.}
  \bibnamefont{Wolynes}}, \bibinfo{journal}{Phys. Rev. B}
  \textbf{\bibinfo{volume}{36}}, \bibinfo{pages}{8552} (\bibinfo{year}{1987}).

\bibitem[{\citenamefont{Goldstein}(1969)}]{Goldstein69JCP}
\bibinfo{author}{\bibfnamefont{M.}~\bibnamefont{Goldstein}},
  \bibinfo{journal}{J. Chem. Phys.} \textbf{\bibinfo{volume}{51}},
  \bibinfo{pages}{3728} (\bibinfo{year}{1969}).

\bibitem[{\citenamefont{Franz and Parisi}(1997)}]{PhysRevLett.79.2486}
\bibinfo{author}{\bibfnamefont{S.}~\bibnamefont{Franz}} \bibnamefont{and}
  \bibinfo{author}{\bibfnamefont{G.}~\bibnamefont{Parisi}},
  \bibinfo{journal}{Phys. Rev. Lett.} \textbf{\bibinfo{volume}{79}},
  \bibinfo{pages}{2486} (\bibinfo{year}{1997}).

\bibitem[{\citenamefont{Franz and Parisi}(2013)}]{Franz_2013}
\bibinfo{author}{\bibfnamefont{S.}~\bibnamefont{Franz}} \bibnamefont{and}
  \bibinfo{author}{\bibfnamefont{G.}~\bibnamefont{Parisi}},
  \bibinfo{journal}{J. Stat. Mech.} \textbf{\bibinfo{volume}{2013}},
  \bibinfo{pages}{P11012} (\bibinfo{year}{2013}).

\bibitem[{\citenamefont{Berthier}(2013)}]{PhysRevE.88.022313}
\bibinfo{author}{\bibfnamefont{L.}~\bibnamefont{Berthier}},
  \bibinfo{journal}{Phys. Rev. E} \textbf{\bibinfo{volume}{88}},
  \bibinfo{pages}{022313} (\bibinfo{year}{2013}).

\bibitem[{\citenamefont{Parisi and Seoane}(2014)}]{PhysRevE.89.022309}
\bibinfo{author}{\bibfnamefont{G.}~\bibnamefont{Parisi}} \bibnamefont{and}
  \bibinfo{author}{\bibfnamefont{B.}~\bibnamefont{Seoane}},
  \bibinfo{journal}{Phys. Rev. E} \textbf{\bibinfo{volume}{89}},
  \bibinfo{pages}{022309} (\bibinfo{year}{2014}).

\bibitem[{\citenamefont{Berthier and Jack}(2015)}]{PhysRevLett.114.205701}
\bibinfo{author}{\bibfnamefont{L.}~\bibnamefont{Berthier}} \bibnamefont{and}
  \bibinfo{author}{\bibfnamefont{R.~L.} \bibnamefont{Jack}},
  \bibinfo{journal}{Phys. Rev. Lett.} \textbf{\bibinfo{volume}{114}},
  \bibinfo{pages}{205701} (\bibinfo{year}{2015}).

\bibitem[{\citenamefont{Berthier and Coslovich}(2014)}]{Berthier11668}
\bibinfo{author}{\bibfnamefont{L.}~\bibnamefont{Berthier}} \bibnamefont{and}
  \bibinfo{author}{\bibfnamefont{D.}~\bibnamefont{Coslovich}},
  \bibinfo{journal}{Proc. Natl. Acad. Sci. USA} \textbf{\bibinfo{volume}{111}},
  \bibinfo{pages}{11668} (\bibinfo{year}{2014}).

\bibitem[{\citenamefont{Berthier et~al.}(2017)\citenamefont{Berthier,
  Charbonneau, Coslovich, Ninarello, Ozawa, and Yaida}}]{Berthier11356}
\bibinfo{author}{\bibfnamefont{L.}~\bibnamefont{Berthier}},
  \bibinfo{author}{\bibfnamefont{P.}~\bibnamefont{Charbonneau}},
  \bibinfo{author}{\bibfnamefont{D.}~\bibnamefont{Coslovich}},
  \bibinfo{author}{\bibfnamefont{A.}~\bibnamefont{Ninarello}},
  \bibinfo{author}{\bibfnamefont{M.}~\bibnamefont{Ozawa}}, \bibnamefont{and}
  \bibinfo{author}{\bibfnamefont{S.}~\bibnamefont{Yaida}},
  \bibinfo{journal}{Proc. Natl. Acad. Sci. USA} \textbf{\bibinfo{volume}{114}},
  \bibinfo{pages}{11356} (\bibinfo{year}{2017}).

\bibitem[{\citenamefont{Palmer}(1982)}]{Palmer82AdvPhys}
\bibinfo{author}{\bibfnamefont{R.}~\bibnamefont{Palmer}},
  \bibinfo{journal}{Adv. Phys.} \textbf{\bibinfo{volume}{31}},
  \bibinfo{pages}{669} (\bibinfo{year}{1982}).

\bibitem[{\citenamefont{Kirkpatrick and Thirumalai}(2015)}]{RevModPhys.87.183}
\bibinfo{author}{\bibfnamefont{T.~R.} \bibnamefont{Kirkpatrick}}
  \bibnamefont{and}
  \bibinfo{author}{\bibfnamefont{D.}~\bibnamefont{Thirumalai}},
  \bibinfo{journal}{Rev. Mod. Phys.} \textbf{\bibinfo{volume}{87}},
  \bibinfo{pages}{183} (\bibinfo{year}{2015}).

\bibitem[{\citenamefont{M\'{e}zard and Parisi}(2012)}]{FPbook}
\bibinfo{author}{\bibfnamefont{M.}~\bibnamefont{M\'{e}zard}} \bibnamefont{and}
  \bibinfo{author}{\bibfnamefont{G.}~\bibnamefont{Parisi}},
  \emph{\bibinfo{title}{Glasses and Replicas}} (\bibinfo{publisher}{John Wiley
  \& Sons, Ltd}, \bibinfo{year}{2012}), chap.~\bibinfo{chapter}{4}, pp.
  \bibinfo{pages}{151--191}, ISBN \bibinfo{isbn}{9781118202470}.

\bibitem[{\citenamefont{Berthier et~al.}(2019)\citenamefont{Berthier, Ozawa,
  and Scalliet}}]{Berth2019}
\bibinfo{author}{\bibfnamefont{L.}~\bibnamefont{Berthier}},
  \bibinfo{author}{\bibfnamefont{M.}~\bibnamefont{Ozawa}}, \bibnamefont{and}
  \bibinfo{author}{\bibfnamefont{C.}~\bibnamefont{Scalliet}},
  \bibinfo{journal}{J. Chem. Phys.} \textbf{\bibinfo{volume}{150}},
  \bibinfo{pages}{160902} (\bibinfo{year}{2019}).

\bibitem[{\citenamefont{Kirkpatrick and
  Thirumalai}(1989)}]{Kirkpatrick89JPhysA}
\bibinfo{author}{\bibfnamefont{T.~R.} \bibnamefont{Kirkpatrick}}
  \bibnamefont{and}
  \bibinfo{author}{\bibfnamefont{D.}~\bibnamefont{Thirumalai}},
  \bibinfo{journal}{J. Phys. A} \textbf{\bibinfo{volume}{22}},
  \bibinfo{pages}{L149} (\bibinfo{year}{1989}).

\bibitem[{\citenamefont{Kirkpatrick and Thirumalai}(1988)}]{PhysRevA.37.4439}
\bibinfo{author}{\bibfnamefont{T.~R.} \bibnamefont{Kirkpatrick}}
  \bibnamefont{and}
  \bibinfo{author}{\bibfnamefont{D.}~\bibnamefont{Thirumalai}},
  \bibinfo{journal}{Phys. Rev. A} \textbf{\bibinfo{volume}{37}},
  \bibinfo{pages}{4439} (\bibinfo{year}{1988}).

\bibitem[{\citenamefont{Flenner et~al.}(2011)\citenamefont{Flenner, Zhang, and
  Szamel}}]{Flenner:2011ht}
\bibinfo{author}{\bibfnamefont{E.}~\bibnamefont{Flenner}},
  \bibinfo{author}{\bibfnamefont{M.}~\bibnamefont{Zhang}}, \bibnamefont{and}
  \bibinfo{author}{\bibfnamefont{G.}~\bibnamefont{Szamel}},
  \bibinfo{journal}{Phys. Rev. E} \textbf{\bibinfo{volume}{83}},
  \bibinfo{pages}{051501} (\bibinfo{year}{2011}).

\bibitem[{\citenamefont{Flenner and Szamel}(2010)}]{Flenner:2010io}
\bibinfo{author}{\bibfnamefont{E.}~\bibnamefont{Flenner}} \bibnamefont{and}
  \bibinfo{author}{\bibfnamefont{G.}~\bibnamefont{Szamel}},
  \bibinfo{journal}{Phys. Rev. Lett.} \textbf{\bibinfo{volume}{105}},
  \bibinfo{pages}{217801} (\bibinfo{year}{2010}).

\bibitem[{\citenamefont{Ediger et~al.}(1996)\citenamefont{Ediger, Angell, and
  Nagel}}]{EdigerMD1996}
\bibinfo{author}{\bibfnamefont{M.~D.} \bibnamefont{Ediger}},
  \bibinfo{author}{\bibfnamefont{C.~A.} \bibnamefont{Angell}},
  \bibnamefont{and} \bibinfo{author}{\bibfnamefont{S.~R.} \bibnamefont{Nagel}},
  \bibinfo{journal}{J. Phys. Chem.} \textbf{\bibinfo{volume}{100}},
  \bibinfo{pages}{13200} (\bibinfo{year}{1996}).

\bibitem[{\citenamefont{Ediger}(2000)}]{EdigerMD2000}
\bibinfo{author}{\bibfnamefont{M.~D.} \bibnamefont{Ediger}},
  \bibinfo{journal}{Ann. Rev. Phys. Chem.} \textbf{\bibinfo{volume}{51}},
  \bibinfo{pages}{99} (\bibinfo{year}{2000}).

\bibitem[{\citenamefont{Biroli and Garrahan}(2013)}]{BiroliG2013}
\bibinfo{author}{\bibfnamefont{G.}~\bibnamefont{Biroli}} \bibnamefont{and}
  \bibinfo{author}{\bibfnamefont{J.~P.} \bibnamefont{Garrahan}},
  \bibinfo{journal}{J. Chem. Phys.} \textbf{\bibinfo{volume}{138}},
  \bibinfo{pages}{12A301} (\bibinfo{year}{2013}).

\bibitem[{\citenamefont{Busch et~al.}(1998)\citenamefont{Busch, Liu, and
  Johnson}}]{BuschR1998}
\bibinfo{author}{\bibfnamefont{R.}~\bibnamefont{Busch}},
  \bibinfo{author}{\bibfnamefont{W.}~\bibnamefont{Liu}}, \bibnamefont{and}
  \bibinfo{author}{\bibfnamefont{W.~L.} \bibnamefont{Johnson}},
  \bibinfo{journal}{J. Appl. Phys.} \textbf{\bibinfo{volume}{83}},
  \bibinfo{pages}{4134} (\bibinfo{year}{1998}).

\bibitem[{\citenamefont{Cho et~al.}(2020)\citenamefont{Cho, Mugnai,
  Kirkpatrick, and Thirumalai}}]{cho2019fragiletostrong}
\bibinfo{author}{\bibfnamefont{H.~W.} \bibnamefont{Cho}},
  \bibinfo{author}{\bibfnamefont{M.~L.} \bibnamefont{Mugnai}},
  \bibinfo{author}{\bibfnamefont{T.~R.} \bibnamefont{Kirkpatrick}},
  \bibnamefont{and}
  \bibinfo{author}{\bibfnamefont{D.}~\bibnamefont{Thirumalai}},
  \bibinfo{journal}{Phys. Rev. E} \textbf{\bibinfo{volume}{101}},
  \bibinfo{pages}{032605} (\bibinfo{year}{2020}).

\bibitem[{\citenamefont{Angell}(1995)}]{Angell1924}
\bibinfo{author}{\bibfnamefont{C.~A.} \bibnamefont{Angell}},
  \bibinfo{journal}{Science} \textbf{\bibinfo{volume}{267}},
  \bibinfo{pages}{1924} (\bibinfo{year}{1995}).

\bibitem[{\citenamefont{Kang et~al.}(2015)\citenamefont{Kang, Yoon, Thirumalai,
  and Hyeon}}]{Kang05PRL}
\bibinfo{author}{\bibfnamefont{H.}~\bibnamefont{Kang}},
  \bibinfo{author}{\bibfnamefont{Y.-G.} \bibnamefont{Yoon}},
  \bibinfo{author}{\bibfnamefont{D.}~\bibnamefont{Thirumalai}},
  \bibnamefont{and} \bibinfo{author}{\bibfnamefont{C.}~\bibnamefont{Hyeon}},
  \bibinfo{journal}{Phys. Rev. Lett.} \textbf{\bibinfo{volume}{115}},
  \bibinfo{pages}{198102} (\bibinfo{year}{2015}).

\end{thebibliography}
 
\end{document}


\title{Supplementary Information for\\ ``RFOT theory for glassy dynamics in a single condensed polymer''}

\author{Hyun Woo Cho$^{1}$, Guang Shi$^{1}$, T. R. Kirkpatrick$^{2}$ and D. Thirumalai$^{1}$} 
\affiliation{$^{1}$Department of Chemistry, University of Texas at Austin, Austin, Texas 78712, USA}
\affiliation{$^{2}$Institute for Physical Science and Technology, University of Maryland, College Park, Maryland 20742, USA}

\date{\today}
\maketitle

\section{I. Simulation model and methods}

\subsection{1. Single polymer model}
We consider a single polymer consisting of  $N=128$ beads that are linearly connected by a harmonic potential, $E_{\text{bond}}(r)=\frac{1}{2}k_b(r-r_0)^2$ where $r$ is the distance between two bonded monomers, and the parameters $k_b$ and $r_0$ are taken to be $2,000\epsilon_{LJ} /\sigma^2$ and $0.9/\sigma$, respectively. Pairs of monomers $i$ and $j$ that are separated along the polymer by more than two bonds with a spatial distance, $r_{ij}$, interact via a Lennard-Jones (LJ) potential, $E_{\text{pair}}(r_{ij})=4\epsilon_{LJ} [(\sigma/r_{ij})^{12}-(\sigma/r_{ij})^6]$. We truncate and shift the LJ potential, setting it  to zero at $r_{ij}=2.5\sigma$. In the simulations, $\epsilon_{LJ}$ and $\sigma$ are the units of energy and length, respectively.

\subsection{2. Preparation of equilibrium conformations}

We performed parallel tempering Monte Carlo (PTMC) simulations \cite{FrenkelSmit} to efficiently sample the equilibrium configurations of the polymers. We used $n_r=32$ independent replicas containing a single polymer at temperatures $T_i$ with $T_1<T_2<\cdots<T_{n_r-1}<T_{n_r=32}$. In the PTMC simulations, the positions of the polymers in the replicas are evolved independently. A monomer is randomly chosen and displaced by $\vec{\delta}$ whose components are uniformly sampled from $[-0.05,0.05]$. The trial position is accepted using the standard Metropolis acceptance probability, $p=\min [1,\exp[-(E_n-E_o)/k_BT]]$, where $E_n$ and $E_o$ are the potential energies of the trial and original configurations, respectively, and the Boltzmann constant $k_B$ is set to unity. We define Monte Carlo steps (MCSs) as the total number of attempted translational moves divided by $n_r$, the number of replicas. At every $N_{swap}=10^4$ MCSs, two neighboring replicas $i$ and $i+1$ are randomly selected, and their configurations are swapped with a probability $p=\min[1,\exp[(1/k_BT_i-1/k_BT_{i+1})(E_i-E_{i+1})]]$, where $E_i$ is the potential energy of the $i^{th}$ replica. 

We carried out three sets of the PTMC simulations. In the first set, $T$ ranges from $T_1=0.35$ to $T_{32}=1.125$ with $T_{gap}=0.025$, where $T_{gap}$ is a temperature difference between the neighboring replicas, $T_{gap}=T_{i+1}-T_{i}$. $(T_1,T_{32}; T_{gap})$ in the second and third sets are $(1.2, 2.75;0.05)$ and $(2.8, 4.35;0.05)$, respectively. All sets of the simulations are run for $10^8$ MCSs. 

\subsection{3. Estimation of $\langle Q_{\text{stat}}\rangle$ and $\chi_{\text{stat}}$}

\begin {figure}
\centering
\includegraphics  [width=3.3in] {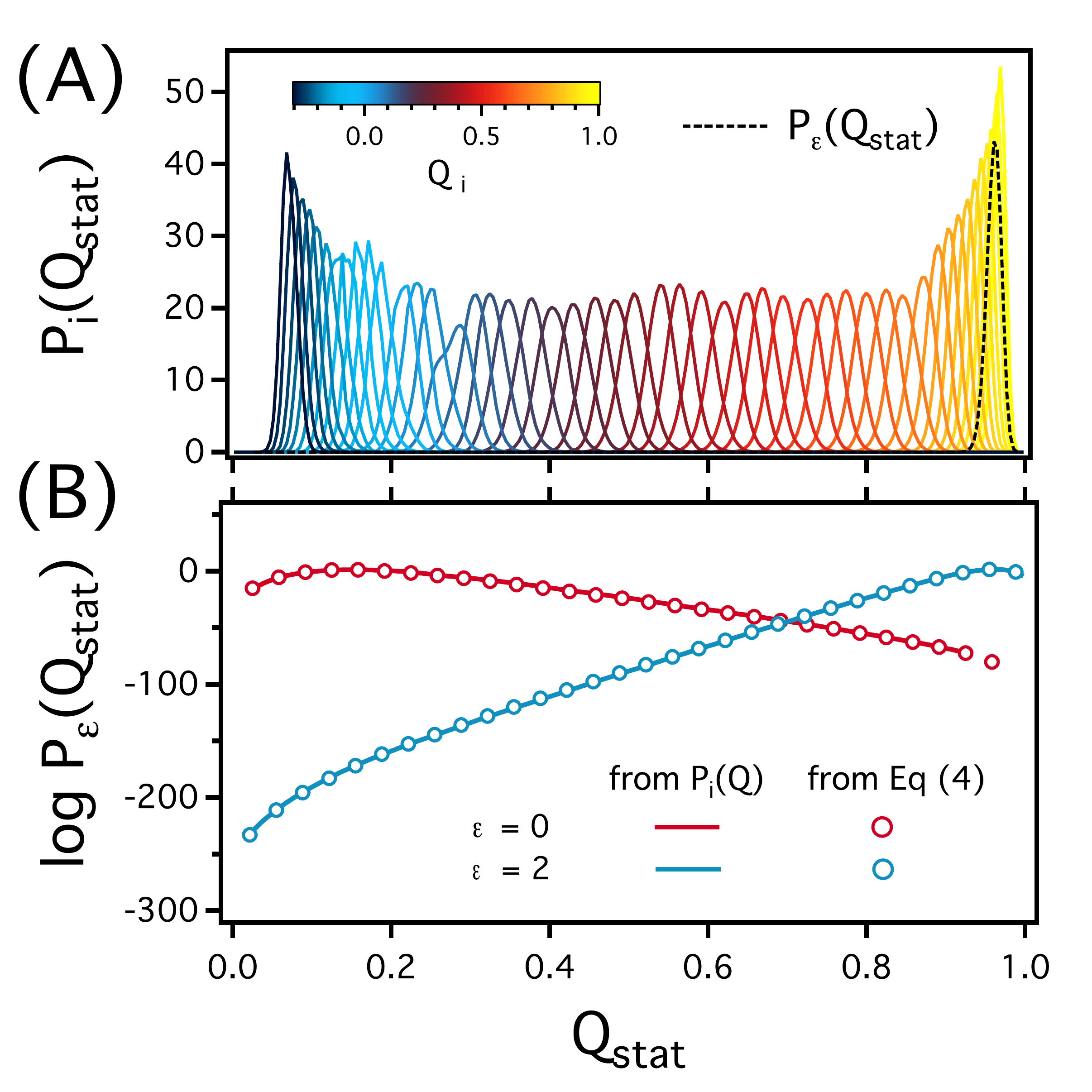}
\caption{(A) Distributions of $Q_{\text{stat}}$, $P_i(Q_{\text{stat}})$, calculated using $n_r=48$ replicas at $T=0.35$ and $\epsilon=2$. The curves are colored according to the value of $Q_i$. $P_\epsilon(Q_{\text{stat}})$ is obtained from unbiased $P_i(Q_{\text{stat}})$ using Eqs~(\ref{Rep1}) and (\ref{Rep2}) (the black dashed curve). (B) The distributions of $Q_{\text{stat}}$ at $T=0.35$ with $\epsilon = 0$ is in red, ) and $\epsilon = 2$ is given by the blue curve. 
Two independent HREMC (Hamiltonian replica exchange MC) simulations were carried out for $\epsilon=0$ and $2$, then $P_\epsilon(Q_{\text{stat}})$ (the solid curves) were obtained by reconstructing $P_i(Q_{\text{stat}})$ as in (A). The open symbols represent $P_\epsilon(Q_{\text{stat}})$ obtained from one another using Eq~(\ref{Rew}). The two methods result in identical $P_\epsilon(Q_{\text{stat}})$.}
\label{replica}
\end{figure}

We calculated $\langle Q_{\text{stat}}\rangle$ and $\chi_{\text{stat}}$ from the distribution $P_\epsilon(Q_{\text{stat}})$ using $\langle Q_{\text{stat}}\rangle=\int_0^1 Q P_\epsilon(Q)dQ$ and $\chi_{\text{stat}}=N(\langle Q_{\text{stat}}^2\rangle-\langle Q_{\text{stat}}\rangle^2)$, where $\langle Q_{\text{stat}}^2\rangle=\int_0^1Q^2 P_\epsilon(Q)dQ$.  In order to obtain $P_\epsilon (Q_{\text{stat}})$, over a broad range of $Q_{\text{stat}}$, we used  umbrella sampling technique while implementing the Hamiltonian replica exchange MC (HREMC) simulations \cite{Berthier11356,PhysRevE.88.022313,Berthier11668}. We used $n_r=48$ replicas, and the energy function $E_i$ of each replica is expressed as,
\begin{eqnarray}
E_i=E_\epsilon(\{\vec{r}\}|\{\vec{r}_0\})+E_{Q_i}(Q_{\text{stat}}),
\label{Hamil}
\end{eqnarray}
where $i$ is the index of the replicas, and $E_{Q_i}(Q_{\text{stat}})$ is a harmonic bias potential expressed as $E_{Q_i}(Q_{\text{stat}})=NW(Q_i-Q_{\text{stat}})^2$, which restricts the distribution of $Q_{\text{stat}}$, $P_i(Q_{\text{stat}})$, near $Q_i$. The position of the single polymer in each replica is evolved via MC simulations for $N_{swap}$ MCSs in  the same way as in the PTMC simulations. At every $N_{swap}$, the configurations of neighboring replicas are swapped with a probability $p=\min[1,\exp[(E_i-E_{i+1})/k_BT]$. The strength of the bias potential, $W$, in this simulation is $4\epsilon_{LJ}$ and $Q_i$'s of each replica are taken for their distributions to cover a full range of $Q_{\text{stat}}$ from 0 to 1 as shown in Figure~\ref{replica} (A). $N_{swap}$ ranges from $10^3$ to $10^4$ MCSs and the simulations are run for $5\times10^6$ to $10^8$ MCSs. Note that $P_\epsilon(Q_{\text{stat}})$ is expressed in terms of $P_i$'s \cite{FrenkelSmit}, 
\begin{eqnarray}
P_\epsilon(Q_{\text{stat}})=\frac{\sum_{i=1}^{n_r}P_i(Q_{\text{stat}})}{\sum_{i=1}^{n_r}g_i \exp[-E_{bias}(Q_i)/k_BT]},
\label{Rep1}
\end{eqnarray} 
where $g_i$ obeys,
\begin{eqnarray}
g_i^{-1}=\int_0^1dQ \frac{\sum_{j=1}^{n_r}P_i(Q)}{\sum_{j=1}^{n_r}g_j \exp[[E_{Q_i}(Q)-E_{Q_j}(Q)]/k_BT]}.
\label{Rep2}
\end{eqnarray} 
The values of $g_i$ were obtained in a self-consistent manner using Eq~(\ref{Rep2}), from which $P_\epsilon(Q_{\text{stat}})$ was determined by Eq~(\ref{Rep1}) (the dashed line in Figure~\ref{replica} (A)). 

Note that for a given set of $\{\vec{r}_0\}$ and $T$, the distributions of $Q_{\text{stat}}$ with two different $\epsilon$ values, say $\epsilon_1$ and $\epsilon_2$, follow a relation,  
\begin{eqnarray}
P_{\epsilon_1}(Q_{\text{stat}})=\frac{P_{\epsilon_2}(Q_{\text{stat}})\exp[-Q_{\text{stat}}\Delta \epsilon/k_BT]}{\int_0^1dQ  P_{\epsilon_2}(Q) \exp[-Q\Delta \epsilon/k_BT] },
\label{Rew}
\end{eqnarray} 
where $\Delta \epsilon$ is $\epsilon_1-\epsilon_2$. Figure~2 confirms the validity of Eq~(\ref{Rew}). First, we evaluate $P_{\epsilon}(Q_{\text{stat}})$ at $\epsilon=0$ and $2$ with same $\{\vec{r}_0\}$ with $T=0.35$ through Eqs~(\ref{Hamil}) to (\ref{Rep2}) (the solid lines in Figure~2). Then, we obtain one from the other using Eq~(\ref{Rew}) (the open circles). As shown in the graph, both the methods result in identical $P_{\epsilon}(Q_{\text{stat}})$. This suggests that $P_{\epsilon}(Q_{\text{stat}})$, over broad ranges of $\epsilon$, can be calculated from a single $P_{\epsilon}(Q_{\text{stat}})$ without carrying out multiple HREMC simulations. Hence, we (1) chose $\{\vec{r}_0\}$ at $T$ among the equilibrium configurations obtained by the PTMC simulation, (2) estimated $P_{\epsilon=0}(Q_{\text{stat}})$ with the HREMC simulations, and (3) determine $P_{\epsilon}(Q_{\text{stat}})$ at various $\epsilon$ from $P_{\epsilon=0}(Q_{\text{stat}})$ using Eq~(\ref{Rew}). The distributions at each $\epsilon$ are averaged over 80 to 112 values of $\{\vec{r}_0\}$. 

\subsection{4. Dynamic Monte Carlo simulation}

In order to investigate the dynamics of a single condensed polymer (SCP), we carried out dynamic Monte Carlo (dMC) simulations, where at every MCS, a randomly chosen monomer was displaced by $\vec{\delta}$ and the resulting configuration was accepted using the standard Metropolis criterion, as in the PTMC simulation. The maximum displacement of each component of $\vec{\delta}$ is 0.01$\sigma$. The MCS (the number of attempted translational moves) normalized by $N$ is used as the unit of time.

\section{II. Conformational phase behavior of single polymers}

\begin {figure}
\centering
\includegraphics  [width=3.3in] {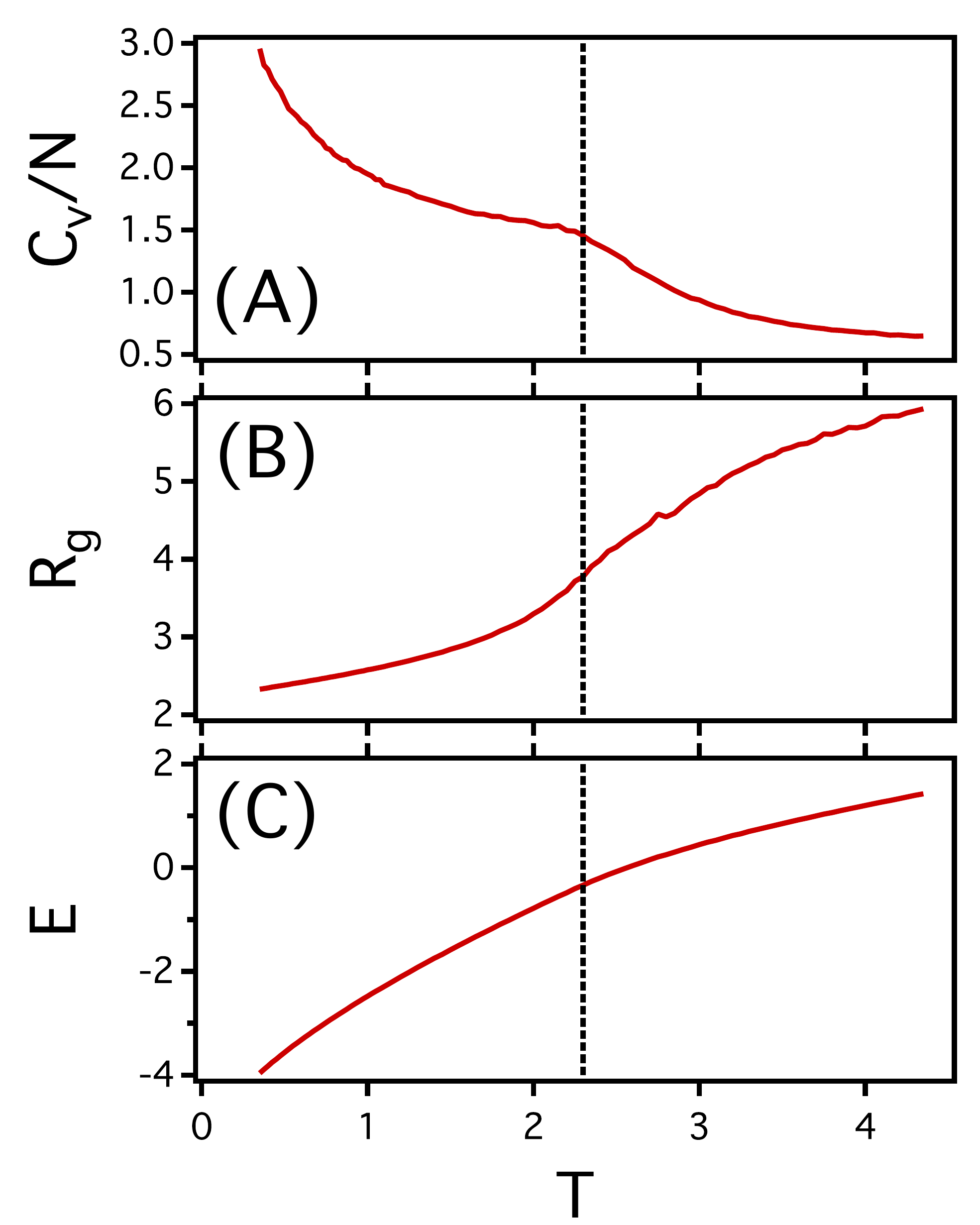}
\caption{(A) $C_v$, (B) $R_g$, and (C) $E$ of the single polymer as a function of $T$. The black dashed line indicates the transition temperature $T_\theta$ where $C_v$ has a shoulder-like structure.}
\label{collapse}
\end{figure}

As $T$ decreases, polymers undergo a coil to globular transition. Such a collapse transition takes place in a continuous manner, therefore the heat capacity $C_v$ of the  polymer has a modest peak or a shoulder-like structure at the transition temperature $T_\theta$ \cite{jcpsingle1,jcpsingle2}, because $N$ is not large. We calculated $C_v$ using  fluctuation in the potential energy $E$, 
\begin{eqnarray}
C_v=\frac{1}{k_BT^2}(\langle E^2 \rangle-\langle E \rangle ^2 ).
\label{Heat}
\end{eqnarray} 
Figure~\ref{collapse} (A) shows that $C_v$ increases with a decrease in $T$ with  a shoulder at $T\simeq2.3$ (the black dashed line). In Figure~\ref{collapse} (B), we plotted the radius gyration $R_g=\sqrt{\frac{1}{N}\sum_1^N (\vec{r_i}^2-\vec{r}_{cm})}$ as a function of $T$, where $\vec{r_i}$ and $\vec{r}_{cm}$ are position vectors of the $i-th$ monomer and the center of mass of the single polymer, respectively. Figure~\ref{collapse} (B) shows that the size of the single polymer changes rather sharply near $T=2.3$, which confirms that the coil-to-globule transition occurs at $T_\theta=2.3$.   

\begin {figure}
\centering
\includegraphics  [width=3.3in] {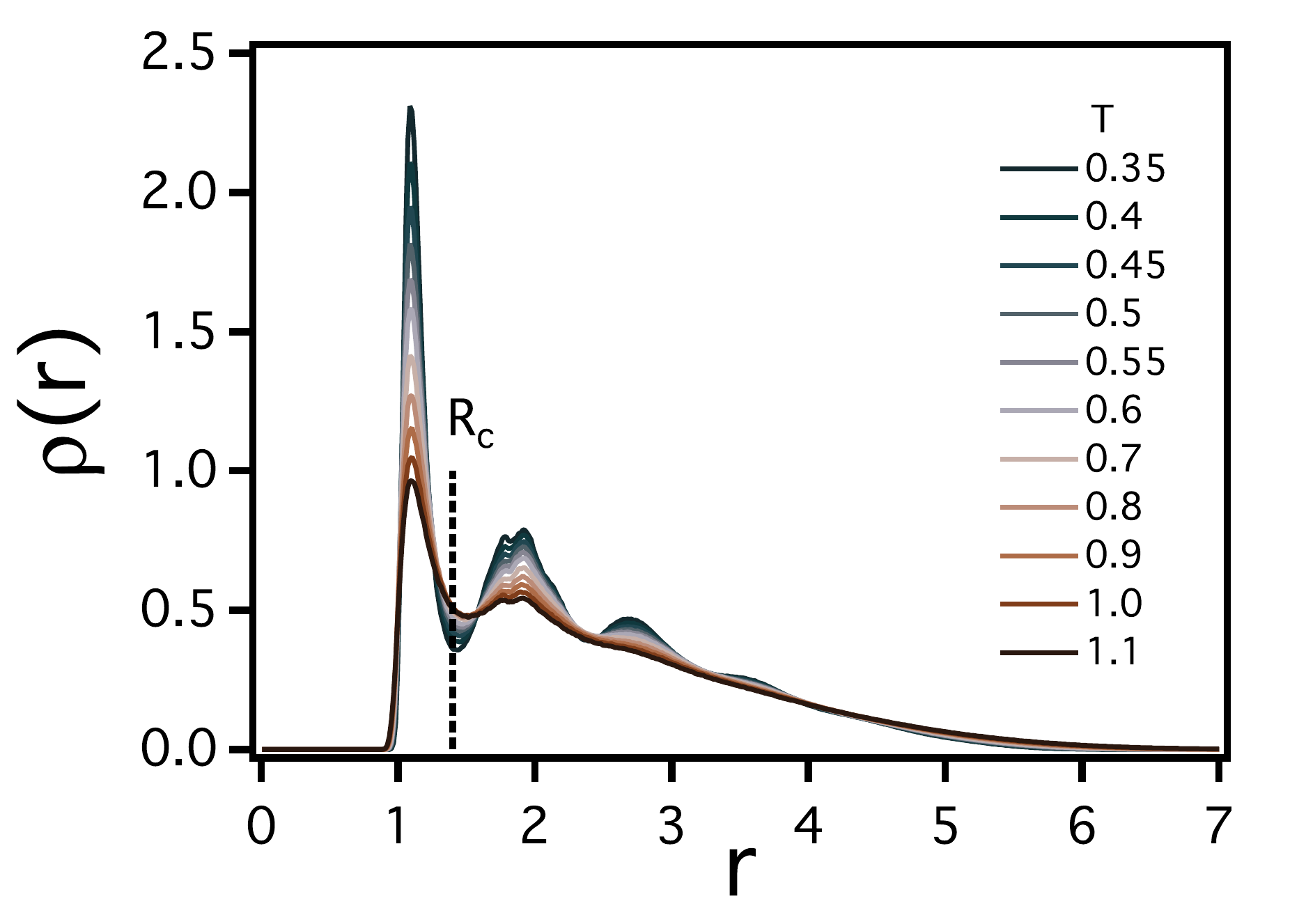}
\caption{The radial distribution function $\rho(r)$ of non-bonded monomers at various $T$. $\rho(r)$ is defined as $\rho(r)=\frac{1}{N_{non}}\sum_{(i,j)'}\delta(|r_i-r_j|-r)$, where $N_{non}$ is the number of non-bonded pairs and $(i,j)'$ indicates that the sum is over all non-bonded pairs of monomers. The discretized peaks, a signature of the ordered spatial arrangement of monomers, are not observed in all range of $T$. $\rho(r)$ has the first minimum at near $R_c=1.4$ (the black dashed line), which is used as a cut-off length when we define a contact between $q_{i,j}$ monomers $i$ and $j$ (see the main text).}
\label{radr}
\end{figure}

When $T$ goes below $T_\theta$, the globular single polymers may become ``crystalline" at the temperature $T_M<T_\theta$. In our single polymer model, however, the crystallization is suppressed in the range of $T$ that we considered ($0.35\leq T\leq 4.35$). The crystallization of the single polymer is a first-order conformational transition, such that $E$ and $R_g$ change abruptly at $T_M$ where $C_v$ has a sharp peak \cite{jcpsingle1}. These interesting features are not observed in Figure~\ref{collapse}. In addition, the radial distribution function $\rho(r)$ of the monomers in Figure 3 confirms again that the spatial arrangement of the monomers is still disordered even at the lowest $T$. 

\section{III. Verification of the metastable states in thermodynamic limit}

\begin {figure*}
\centering
\includegraphics  [width=5in] {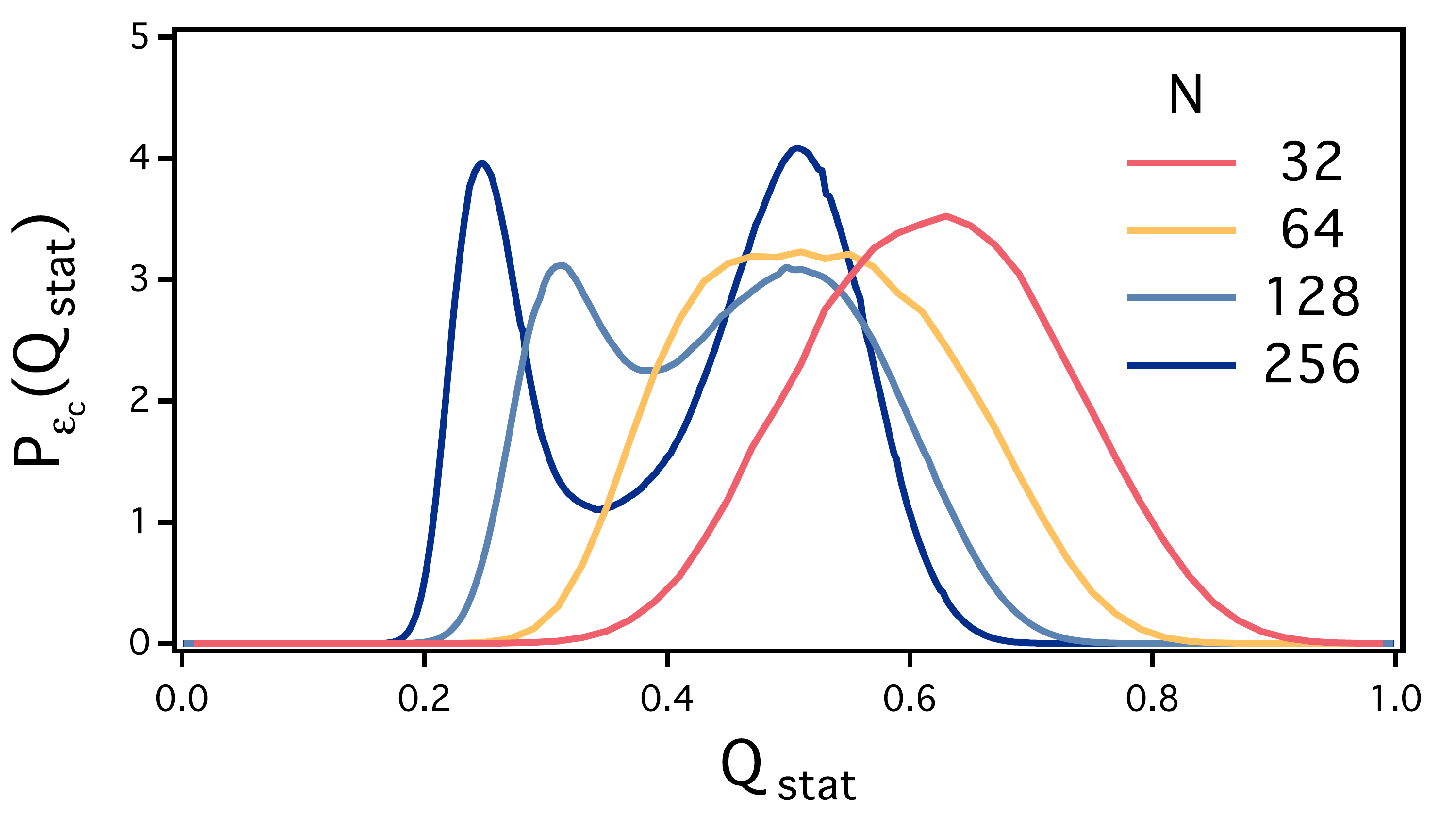}
\caption{$P_{\epsilon_c}(Q_{\text{stat}})$ at $T=0.9$ for various $N$.}
\label{finite}
\end{figure*}

In the main text, we established that the first order transition in $Q_{\text{stat}}$ occurs as $\epsilon (\ne 0)$ varies, proving the existence of the metastable states of the SCP when $N=128$. However, this does not, in principal, guarantee that the metastable states of the SCP should appear in thermodynamic limit ($N\rightarrow\infty$), since we considered only one length of single polymers. In this section, by illustrating that apparent first order transition in $Q_{\text{stat}}$ takes place at larger $N$, we confirm  the existence of the metastable states for larger $N$ as well. It is likely that the transitions become sharper for  $N \gg 1$ although it is difficult to carry out these simulations using the FP method. 

We carried out additional simulations for various $N$ at $T=0.9$ and determined $\epsilon_c$. The distributions of $Q_{\text{stat}}$ at $\epsilon_c$ ($P_{\epsilon_c}(Q_{\text{stat}})$) are depicted in Figure~\ref{finite}. When $N=32$, $P_{\epsilon_c}(Q_{\text{stat}})$ has a single peak, indicating a likely continuous transition in the order parameter, $Q_{\text{stat}}$. As $N$ increases to 256, the peak becomes broader and is split into two peaks. With an increase in $N$, the separation between the two peaks is more pronounced, and the concave character between them becomes deeper. Considering that the negative logarithm of $P_{\epsilon_c}(Q_{\text{stat}})$ is associated with the energy landscape of the transition (Eq (3) of the main text), this means that the metastable states in the SCP are separated by higher energy barriers as $N$ increases, suggesting that the metastable states should exist more stably as $N\rightarrow\infty$.

\section{IV. Determination of $T_d$}

\begin {figure}
\centering
\includegraphics  [width=3.3in] {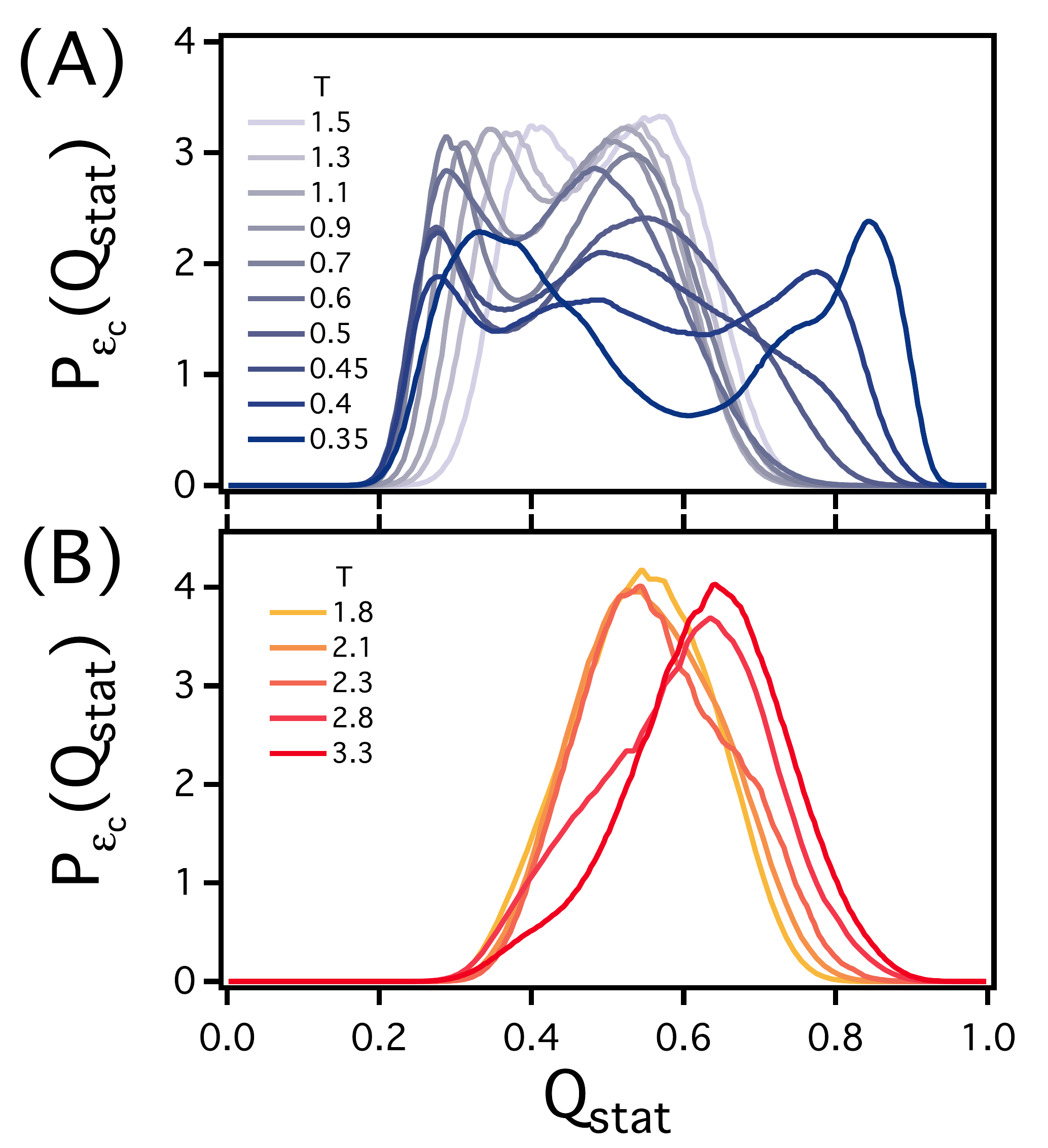}
\caption{$P_{\epsilon_c}(Q_{\text{stat}})$ for (A) $T<1.8$ and (B) $T\leq1.8$.}
\label{transition}
\end{figure}

According to the RFOT theory, the exponential number of metastable states emerge at or below the dynamic transition $T_d$ \cite{PhysRevA.40.1045}, which for finite system would represent a rounded transition. Since in the FP method, the first-order transition characterized by $Q_{\text{stat}}$ at non-zero $\epsilon_c$ is attributed to the existence of the metastable states, the first-order transition nature should disappear and the overlap order parameter should change continuously with $\epsilon$ when $T>T_d$ \cite{PhysRevLett.79.2486,PhysRevLett.114.205701}. Therefore, $T_d$ of the single polymer can be determined as $T$ where the first-oder transition in $Q_{\text{stat}}$ starts to vanish. In Figure~\ref{transition}, we show the distribution of the order parameters at the critical $\epsilon=\epsilon_c$, $P_{\epsilon_c} (Q_{\text{stat}})$, by varying $T$. For $T< 1.8$ (Figure~\ref{transition} (A)), $P_{\epsilon_c}(Q_{\text{stat}})$ is bimodal, which is a signature of the first-order phase transition. On the other hand, when $T$ exceeds 1.8 (Figure~\ref{transition} (B)), $P_{\epsilon_c}(Q_{\text{stat}})$ becomes unimodal, implying that $Q_{\text{stat}}$ changes continuously at $T\geq1.8$. Hence, we determine that $T_d=1.8$. 

\section{V. Random Field Ising Model (RFIM) behavior of single polymers}

\begin {figure}
\centering
\includegraphics  [width=3.3in] {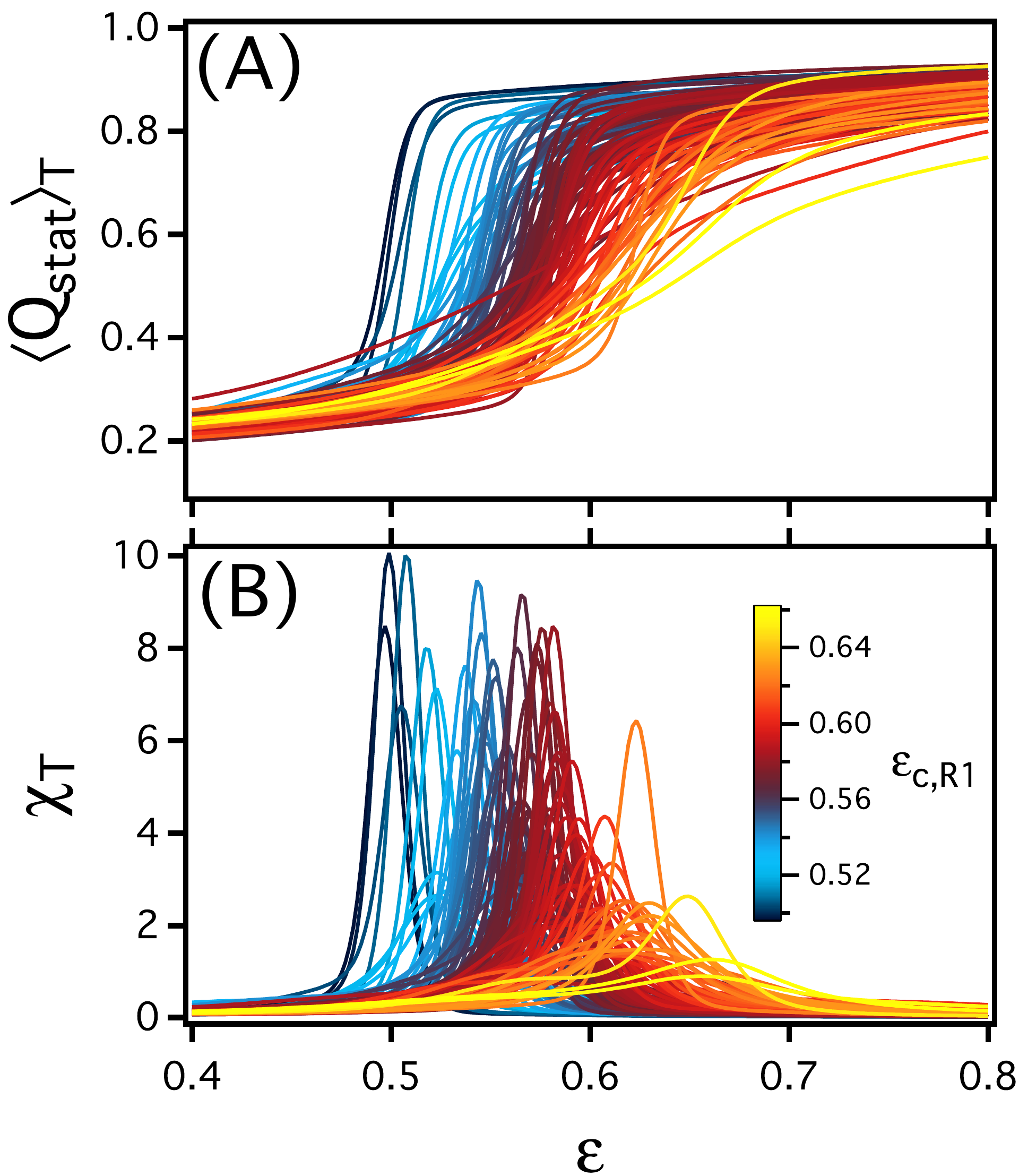}
\caption{(A) The average and (B) the susceptibility of $Q_{\text{stat}}$ as a function of $\epsilon$ for given configurations $\{\vec{r_0}\}$ in Replica 1 at $T=0.35$. Individual curves are colored according to their $\epsilon_c$ (the color bar in (B)), where $\epsilon_c$ is defined as a peak position of $\chi_T$ in the lower panel.}
\label{RFIM1}
\end{figure}

In the FP method for glassy liquids, the external coupling field ($\epsilon$)  introduces  quenched disorder to the configuration $\{\vec{r}\}$ in  Replica 2. The nature of the quenched disorder is determined by the reference configuration $\{\vec{r_0}\}$, which influences the critical fluctuations. Figure~\ref{RFIM1} illustrates the influence of the disorder on the critical behavior of the order parameters. In Figure~\ref{RFIM1} (A), we considered 112 $\{\vec{r_0}\}$ at $T=0.35$ and the estimated the averages of $\langle Q_{\text{stat}}\rangle_T$ with respect to $\epsilon$ for individual $\{\vec{r_0}\}$, where $\langle\cdots\rangle_T$ indicates the thermal average of the property for fixed $\epsilon$ and $\{\vec{r_0}\}$. We plot the corresponding susceptibilities $\chi_T$ in Figure~\ref{RFIM1} (B), where $\chi_T=N(\langle Q_{\text{stat}}^2\rangle_T-\langle Q_{\text{stat}}\rangle_T^2)$. The figure shows that the transitions in $Q_{\text{stat}}$ for various $\{\vec{r_0}\}$ take place at different values of $\epsilon$ with different extent of sharpness, corresponding to peak positions and heights of their $\chi_T$. The contribution of the disorder to the critical behavior of $Q_{\text{stat}}$ is quantified by decomposing the fluctuation $\chi_{\text{stat}}$ into the thermal ($\chi_T$) and disorder ($\chi_D$) contributions \cite{PhysRevLett.114.205701}, where $\chi_D$ is defined as 
\begin{eqnarray}
\chi_D=N\big(\big\langle  \langle Q_{\text{stat}}\rangle_T^2\big\rangle_D - \big\langle \langle Q_{\text{stat}}\rangle_T\big\rangle_D^2 \big),
\label{dis}
\end{eqnarray}  
and $\langle\cdots\rangle_D$ denotes an average over $\{\vec{r_0}\}$ at a given $T$ and $\epsilon$. By definition, the sum of $\chi_D$ and $\chi_T$ is $\chi_{\text{stat}} =\chi_D + \chi_T$.

\begin {figure}
\centering
\includegraphics  [width=3.3in] {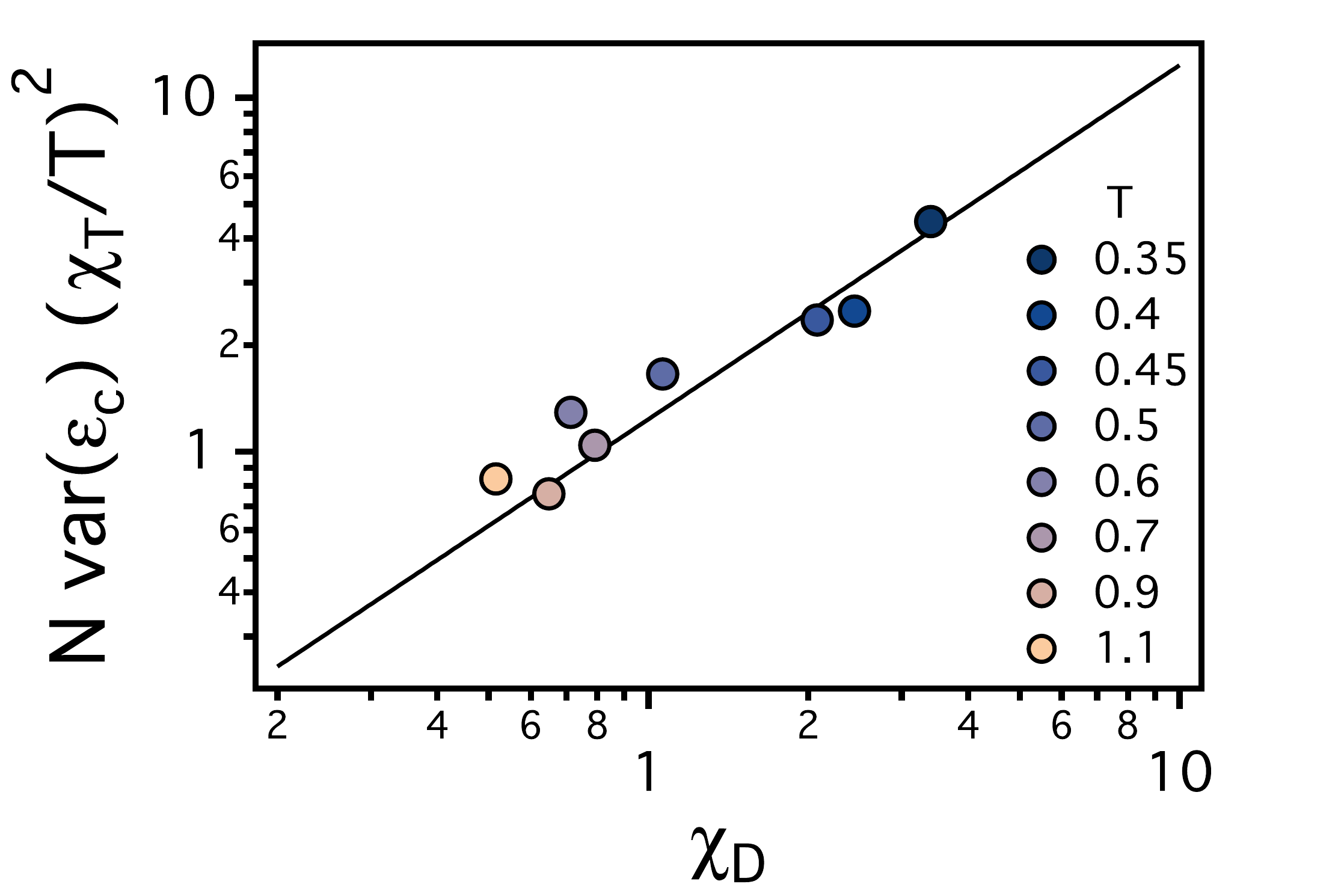}
\caption{Quantitative relation between $\chi_T$ and $\chi_D$. The linear line ($y=1.23x$) is obtained by curve fitting, which confirms the validity of Eq~(\ref{critical})}
\label{RFIM2}
\end{figure}

In glassy liquids, $\chi_D$ grows more rapidly than $\chi_T$ as $T$ decreases. Quantitative relation between $\chi_D$ and $\chi_T$ is expressed as  
\begin{eqnarray}
\chi_D\simeq N\text{var}(\epsilon_c)\Big(\frac{\chi_T}{T}\Big)^2,
\label{critical}
\end{eqnarray}  
where $\text{var}(\epsilon_c)$ is the variance of the critical points among $\{\vec{r_0}\}$ \cite{Berthier1797,berthiercritical1,berthiercritical2}. In Figure~\ref{RFIM2}, we plot $N\text{var}(\epsilon_c)\big(\frac{\chi_T}{T}\big)^2$ with respect to $\chi_D$ for various $T$ (the colored symbols), where $\epsilon_c$ of $\{\vec{r_0}\}$ is determined as a peak position of $\chi_T$. The black solid line ($y\sim x$) confirms the validity of Eq~(\ref{critical}), showing hat $\chi_D$ and $\chi_T$ follows well Eq~(\ref{critical}). This implies that as $T$ approaches $T_K$, the disorder fluctuation of the single polymer makes the dominant contribution to the total fluctuations.  This behavior is reminiscent of the phase transition behavior in the random field Ising model (RFIM)  \cite{PhysRevLett.56.416}. Therefore, we conclude that freezing in single polymeric glass when $\epsilon \ne 0$ could be in  the same universality class as RFIM. The similarity between the RFIM-like behavior and that found in glass forming systems \cite{Berthier1797,berthiercritical1,berthiercritical2} further fortifies the conclusions reached in the main text that the low temperature properties of the SCP, and other compact polymers such as chromosomes, ought to exhibit glass-like behavior in the condensed state. We will be remiss if we did not add that  at $T_K \sim T_c$ the transition has the character of the random first order transition at zero coupling strength.

\bibliography{reference_si}
\bibliographystyle{apsrev}